\documentclass[pra,twocolumn,floatfix,superscriptaddress,longbibliography,notitlepage]{revtex4-1}

\usepackage{graphicx, color, graphpap}
\usepackage{enumitem}
\usepackage{amssymb}
\usepackage{amsthm}
\usepackage{multirow}
\usepackage[colorlinks=true,citecolor=blue,linkcolor=magenta]{hyperref}
\usepackage[T1]{fontenc}
\usepackage{bbm}
\usepackage{thmtools,thm-restate}
\usepackage{verbatim}
\usepackage{mathtools}
\usepackage{titlesec}
\usepackage{amsmath}
\usepackage[title]{appendix}
\usepackage[linesnumbered,ruled,vlined]{algorithm2e}
\SetKwInput{kwInit}{Init}

\usepackage[caption = false]{subfig}

\long\def\ca#1\cb{} 

\newcommand{\braket}[2]{\langle #1 \hspace{1pt} | \hspace{1pt} #2 \rangle}

\newcommand{\ket}[1]{|#1\rangle}               
\newcommand{\bra}[1]{\langle #1|}              

\DeclareMathOperator*{\argmin}{arg\,min}
\renewcommand{\vec}[1]{\boldsymbol{#1}}  

\newcommand{\ad}{^\dagger}

\newcommand*{\id}{\openone}

\newcommand{\thv}{\vec{\theta}}

{}
{}
\newtheorem{theorem}{Theorem}

\newtheorem{corollary}{Corollary}

\newenvironment{specialproof}{\textit{Proof:}}{\hfill$\square$}

\newcommand\Z{\text{Z}}

\usepackage{amssymb}
\usepackage{dsfont}
\usepackage{bbold}

\begin{document}

\title{Quantum Algorithms for Geologic Fracture Networks}

\author{Jessie M. Henderson}
\affiliation{Los Alamos National Laboratory, Los Alamos, New Mexico 87545, USA}
\affiliation{Southern Methodist University, Dallas, Texas 75205, USA}

\author{Marianna Podzorova}
\affiliation{Los Alamos National Laboratory, Los Alamos, New Mexico 87545, USA}
\affiliation{Theoretical Division, Los Alamos National Laboratory, Los Alamos, New Mexico 87545, USA}
\affiliation{The University of Maryland Department of Computer Science and Joint Center for Quantum Information \& Computer Science, College Park, Maryland 20742, USA}

\author{M. Cerezo}
\affiliation{Los Alamos National Laboratory, Los Alamos, New Mexico 87545, USA}
\affiliation{Information Sciences, Los Alamos National Laboratory, Los Alamos, NM 87545, USA}
\affiliation{Center for Nonlinear Studies, Los Alamos National Laboratory, Los Alamos, New Mexico 87544}

\author{John K. Golden}
\affiliation{Los Alamos National Laboratory, Los Alamos, New Mexico 87545, USA}

\author{Leonard Gleyzer}
\affiliation{Los Alamos National Laboratory, Los Alamos, New Mexico 87545, USA}
\affiliation{Brown University, Providence, Rhode Island 02912, USA}

\author{Hari S. Viswanathan}
\affiliation{Los Alamos National Laboratory, Los Alamos, New Mexico 87545, USA}

\author{Daniel O'Malley}
\affiliation{Los Alamos National Laboratory, Los Alamos, New Mexico 87545, USA}

\begin{abstract}
Solving large systems of equations is a challenge for modeling natural phenomena, such as simulating subsurface flow.
To avoid systems that are intractable on current computers, it is often necessary to neglect information at small scales, an approach known as coarse-graining. 
For many practical applications, such as flow in porous, homogenous materials, coarse-graining offers a sufficiently-accurate approximation of the solution.
Unfortunately, fractured systems cannot be accurately coarse-grained, as critical network topology exists at the smallest scales, including topology that can push the network across a percolation threshold.
Therefore, new techniques are necessary to accurately model important fracture systems.
Quantum algorithms for solving linear systems offer a theoretically-exponential improvement over their classical counterparts, and in this work we introduce two quantum algorithms for fractured flow.
The first algorithm, designed for future quantum computers which operate without error, has enormous potential, but we demonstrate that current hardware is too noisy for adequate performance.
The second algorithm, designed to be noise resilient, already performs well for problems of small to medium size (order 10 to 1000 nodes), which we demonstrate experimentally and explain theoretically.
We expect further improvements by leveraging quantum error mitigation and preconditioning.
\end{abstract}

\maketitle

\section{Introduction}\label{sec:introduction}
Simulating geologic flow requires solving systems of equations, a process that can become computationally prohibitive as system dimension increases~\cite{driscoll2017fundamentals}.
Classical computers can thus solve large systems only when information is removed from consideration.  Coarse-graining is one technique for reducing system size.
Originally developed to model multi-scale biochemical systems, it has become an oft-used means of simplifying linear systems, including in the geosciences~\cite{levitt1975computer, warshel1976theoretical,levitt2014birth}.
Specifically, the coarse-graining technique of upscaling can be accurately applied to geological problems involving spatially-large, materially-homogeneous regions.
The technique combines mesh nodes and assigns them an averaged, or upscaled, permeability or other geological feature, losing mesh resolution but, in this context, still preserving approximately accurate solutions~\cite{durlofsky2005upscaling}.

Unfortunately, fracture network problems cannot be classically solved in their entirety, nor can they be accurately solved with upscaling, making simulation of fracture systems one of the most challenging problems in geophysics~\cite{fountain2005fractures,davies1999role,viswanathan2022fluid,laubach2019role}.
Fractures exist over a range of at least $10^{-6}$ to $10^4$ meters, and the computational requirements involved in completely solving systems comprising over ten orders of magnitude quickly become prohibitive.
Such systems also cannot be accurately upscaled because the information thereby lost pertains to small fractures that ought not generally be neglected.
Collectively, such fractures can radically transform the network topology, including by possibly pushing the network over a percolation threshold.
The small fractures can collectively contribute a significant amount of surface area, enabling stronger interaction between the fractures and the rock matrix, potentially providing complete connectivity that would not otherwise exist in a region~\cite{o2016does}.

\begin{figure*}[t!]
\centerline{\includegraphics[width=\linewidth]{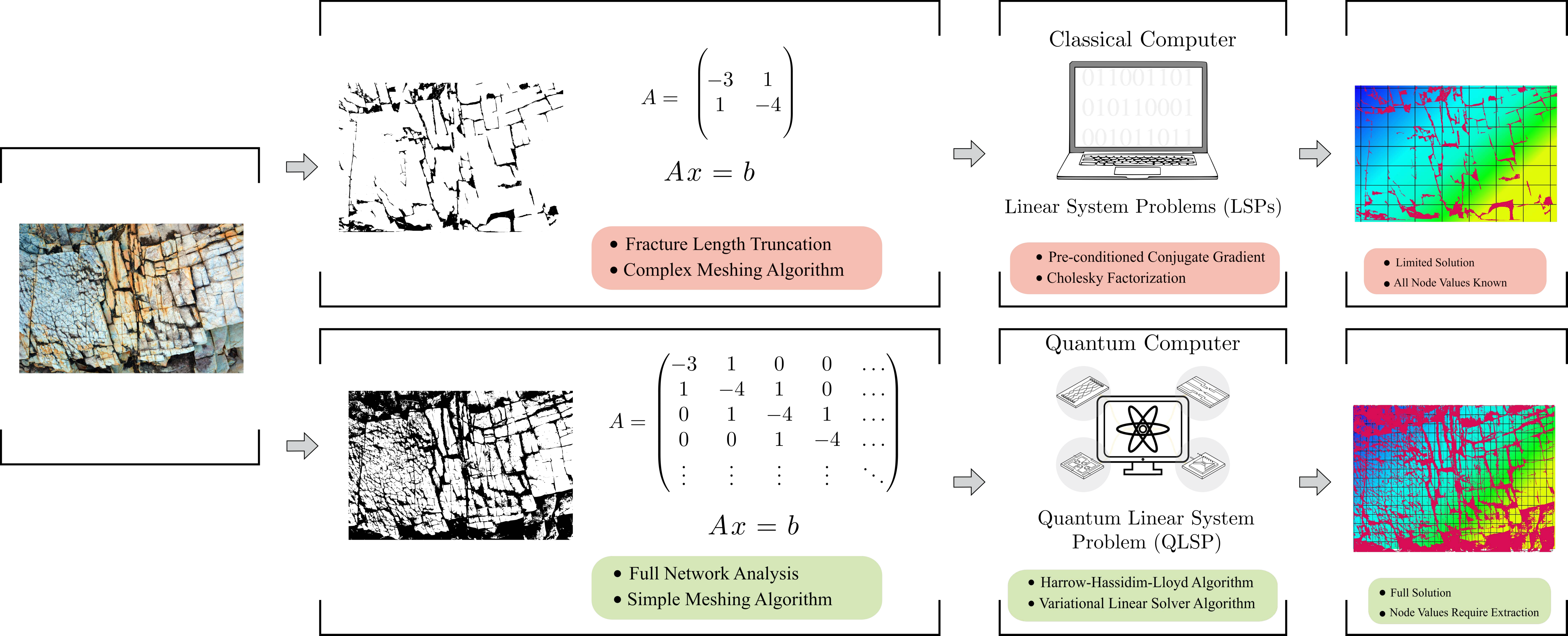}}
\caption{\textbf{Schematic workflow for applying classical and quantum algorithms to fracture flow problems.}  Discretizing fracture systems on classical computers involves reducing the computational cost by truncating the fracture network to exclude small fractures.  This setting-aside of information provides a solution that does not accurately reflect all flow.  Conversely, quantum computing has the potential to solve large, complete fracture systems given properties of quantum mechanics and algorithms designed to take advantage of those properties.}
\label{fig:VLS_for_Geosciences_Workflow}
\end{figure*}

Thus, accurate geologic flow models should include fractures at the entire range of scales.
While advanced meshing techniques~\cite{hyman2015dfnworks} and high-performance simulators~\cite{mills2007simulating} allow inclusion of increased fracture range, even such sophisticated approaches do not make it possible to model the full fracture scale.
So, as illustrated in Fig.~\ref{fig:VLS_for_Geosciences_Workflow}, classical approaches to geologic fracture problems depend upon upscaling that neglects information which can dramatically affect the solution.

By contrast, quantum algorithms provide efficient solutions for solving large linear systems that could include the entire scale of geologic fractures~\cite{harrow2009quantum}.
Properties of quantum computing are fundamentally different than classical counterparts, theoretically permitting the solution of classically intractable problems~\cite{google2019supremacy,zhong2020quantum, nielsen2000quantum}.
Among other benefits, quantum computers store solutions as a vector, $\psi$, containing $2^n$ elements, where $n$ is the number of qubits (or quantum bits).
A quantum computer can thereby solve vast systems of equations with a relatively small number of qubits: $n$ qubits allows for solving a system with $2^n$ variables.
Consider a straightforward example involving a cubic fracture domain comprising one-kilometer and employing a one-centimeter resolution.
Given $10^5$ centimeters to a kilometer, simulating this region would require $(10^5)^3=10^{15}$ nodes.
While a classical computer would thus require $O(10^{15})$ bits, a quantum computer would require only $O(log_2(10^{15}))\approx O(10^1)$ qubits.

This article illustrates using quantum algorithms to solve fracture flow linear systems problems (LSPs) for which upscaling is not appropriate.
We introduce two algorithms and provide proof-of-concept application using IBM’s suite of quantum devices.
We consider problems formulated as a numerical discretization of $\nabla \cdot (k \nabla h) = f$, where $k$ is the permeability, $f$ is a fluid source or sink, and $h$ is the pressure to be computed.  This discretization results in a linear system of equations $(A\textbf{x}=\textbf{b})$, where $A$ is a matrix, and $\textbf{x}$ and $\textbf{b}$ are vectors.  The solution, $\textbf{x}$, represents the pressure at each of the discretized nodes, and quantum algorithms prepare a normalized vector proportional to this solution.

We note that obtaining this solution from a quantum computer works differently than from a classical machine.
Upon quantum algorithm completion, the entire solution is not readily available, and indeed, requires exponential time to obtain~\cite{montanaro2016quantum}.
This is no issue for applications in which the goal is not to \textit{know} the entire solution, but is instead to completely \textit{solve} the problem, such that any portion of the solution that a user obtains is accurate.
Fortunately, fracture networks present just such a situation; ordinarily, we are interested in the pressure at a small, fixed number of nodes on the computational mesh, such as the nodes corresponding to a well location.
Rather than extract the pressure at all nodes from the quantum computer, we need only obtain the pressures at nodes corresponding to the area of interest.
Furthermore, fracture flow problems can be specified in such a way that the complexity required to obtain information about multiple nodes' pressures is reduced.
A procedure that we term `smart encoding' allows obtaining the aggregated pressures of a series of nodes at the computational cost of a single node.
(See Sec.~\ref{sec:Smart_Encoding} online for further details.)

The paper proceeds as follows. Sec.~\ref{sec:fault_tolerant_algs} first presents two algorithms--the Harrow-Hassidim-Lloyd and Subasi-Somma-Orsucci algorithms---that have proven potential for solving LSPs on error-corrected, or fault-tolerant,  quantum computers~\cite{harrow2009quantum}.
Despite the potential for exponential gain in certain cases, the high noise levels of current hardware result in poor performance~\cite{gambetta2020ibmsroadmap, biercuk2021quantum,ai2021exponential,wang2020noise}.
Sec.~\ref{sec:NISQ_algs} then turns to algorithms designed for contemporary, noisy intermediate-scale quantum (NISQ) computers~\cite{peruzzo2014variational,cerezo2020variationalreview,preskill2018quantum,bharti2021noisy}.
Specifically, we experimentally illustrate the noise resilience of the Variational Linear Solver algorithm~\cite{bravo2020variational}, which provides improved solution accuracy even on available error-prone machines for fracture LSPs of small to medium size (10 to 1000 nodes).
We conclude by situating our results and suggesting future improvements.

\section{Results}
\subsection{Algorithms for the Fault-Tolerant Era}\label{sec:fault_tolerant_algs}
The first algorithm for solving quantum linear systems problems (QLSPs) was introduced by Harrow, Hassidim, and Lloyd (HHL)~\cite{harrow2009quantum}.  It solves the sparse $N$-variable system $A\textbf{x}=\textbf{b}$ with a computational complexity that scales polynomially with $\log(N)$ and the condition number, $\kappa$, of the matrix $A$~\cite{harrow2009quantum}.
This provides an exponential speedup over the best classical approaches when $\kappa$ is small, such as when an effective preconditioner is used.
However, the quantum circuit requirements of HHL---when applied to problems of even moderate size---are well-beyond the capabilities of currently available quantum hardware~\cite{scherer2017concrete}. 
This is largely because HHL utilizes complex subroutines, such as Quantum Phase Estimation, which require qubits that operate with almost no quantum noise or error.
On NISQ hardware, HHL is thus impractical for systems of interest; the largest system solved to date using HHL is of dimension $16 \times 16$~\cite{zheng2017solving, lee2019hybrid, pan2014experimental, cai2013experimental, barz2014two, wen2019experimental}.
Once large fault-tolerant quantum computers are developed, the exponential speedup offered by HHL (and variations/improvements thereon) could play a critical role in advancing subsurface flow modeling.

In the interim, progress in QLSP algorithms has occurred in two directions.
The first is to tailor QLSP algorithms to the strengths and weaknesses of current NISQ computers, such as the algorithm we present in Sec.~\ref{sec:NISQ_algs} does~\cite{bravo2020variational,huang2019near,xu2019variational,lee2019hybrid,chen2019hybrid}.
The second is to design algorithms that are still intended for fault-tolerant computers, but which do not rely on as many complex subroutines as HHL and thus may perform adequately on NISQ devices~\cite{subacsi2019quantum,childs2017quantum,chakraborty2019power,wossnig2018quantum}.
One such example is the adiabatic approach of Subasi, Somma, and Orsucci (SSO)~\cite{subacsi2019quantum}. 
This approach requires only a single subroutine, known as Hamiltonian simulation, while still offering the equivalent quantum speed-up of HHL.

\begin{figure}[t]
\centerline{\includegraphics[width=\linewidth]{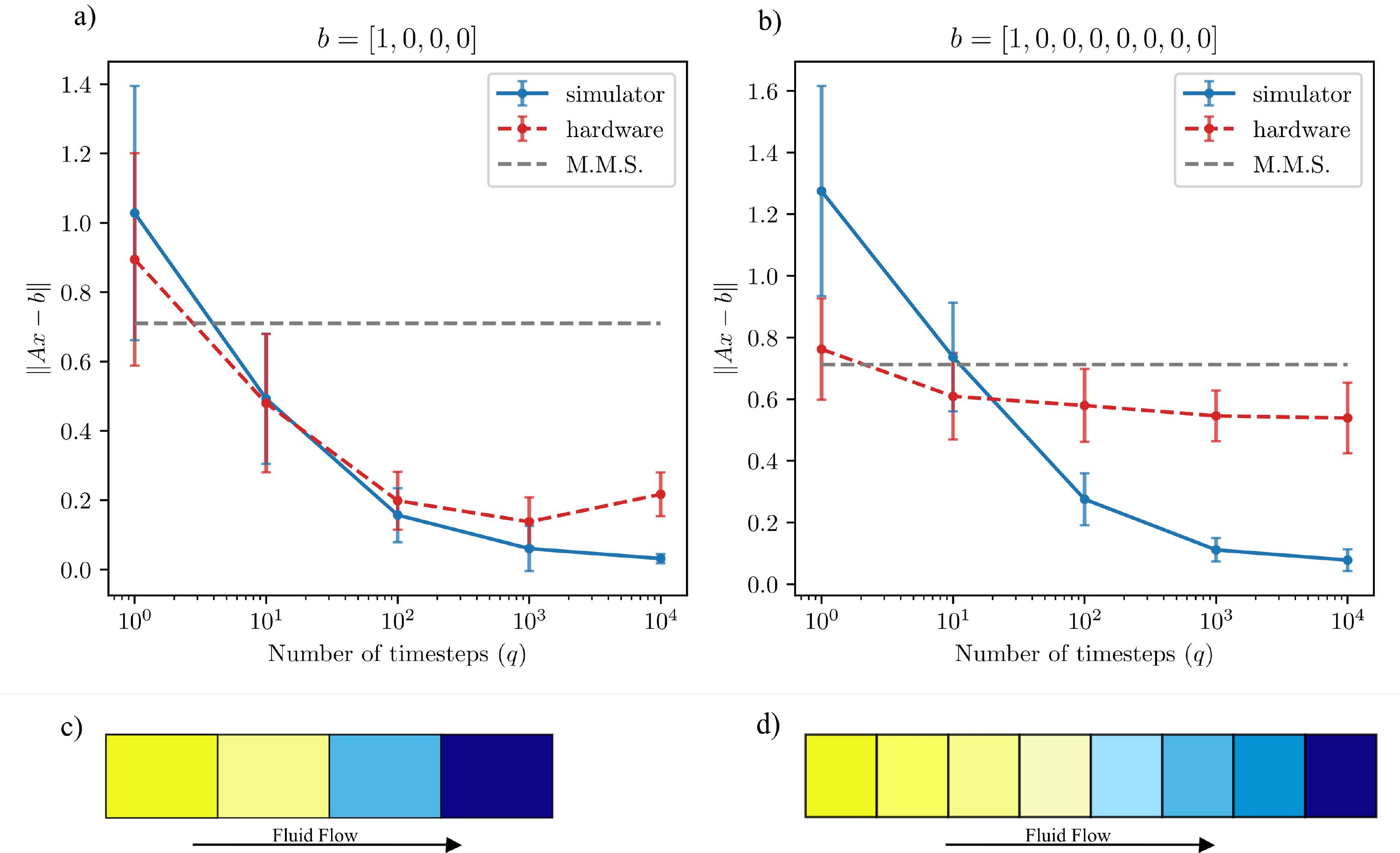}}
\caption{\textbf{Solving 4-node and 8-node fracture systems using the SSO algorithm.}  Subfigure (a) presents results for a one-dimensional grid of 4 nodes, while subfigure (b) does the same for a grid of 8 nodes.  Each subfigure presents the maximum, minimum, and average error in 75 runs on either IBM’s quantum simulator (solid blue line) or the \texttt{ibmq\_rome} quantum computer (dashed red line).  The error is plotted against the number of iterations, $q$, used.  The figures also include a dotted gray line illustrating the error that would be achieved if the returned `solution' from the quantum computer was the result of the maximally-mixed state.  As described in the main text, this serves as a benchmark for assessing the quality of the solution SSO achieved. Subfigures (c) and (d) present cartoons of the problems solved: each color indicates a unique node pressure, with fixed pressures of 1 and 0 on the boundaries.}
\label{fig:SSO_results_main_text}
\end{figure}

Before embarking on the purely NISQ-oriented approach of Sec.~\ref{sec:NISQ_algs}, we tested the SSO algorithm on a collection of very simple subsurface flow problems to assess how well current hardware could handle one fault-tolerant algorithm.
As described in Sec.~\ref{sec:introduction}, the problem was to compute pressures of a one-dimensional grid of either $N=4$ or $N=8$ nodes.
(See subfigures (c) and (d) of Fig.~\ref{fig:SSO_results_main_text} for a cartoon visualization.)
Pressures on the boundaries were fixed, and the answer to the QLSP encoded the internal pressures.

The computational complexity and resulting accuracy of the SSO algorithm depend upon a unitless, user-defined parameter, $q$, which is connected to how long the algorithm is allowed to run. 
We showed that---up to a point---the algorithm returned better results as $q$ increased; for both $N=4$ and $N=8$ problems running on a noiseless quantum simulator, the error $\|A\textbf{x}-\textbf{b}\|$ approached 0 for $q=10^4$.
On the quantum hardware, the average error after an equivalent time was approximately $0.21$ for an $N=4$ problem and $0.54$ for an $N=8$ problem.
(Note that for these problems, $\|\textbf{b}\|=1$.)

Fig.~\ref{fig:SSO_results_main_text} illustrates these results and two noteworthy points.
First, the $N=8$ problem exhibited a clear limit to how much the hardware results would improve with increasing $q$.
Indeed, despite increasing $q$ by four orders of magnitude, the average error when run on the quantum hardware decreased by only about 0.2.
This suggests that---on NISQ-era devices---SSO's utility is limited even for problems with as few as 8 nodes.

Second, Fig.~\ref{fig:SSO_results_main_text} compares the errors achieved on quantum simulators and hardware to the error when obtaining a result from the a quantum state known as the \textit{maximally-mixed state}. 
This comparison contextualizes the quality of the errors achieved by SSO, because the maximally-mixed state corresponds to a state where noise has destroyed all information in the quantum system, and thus can be characterized as  one of random information.
Specifically, for the fracture flow LSPs we solved, obtaining a result from a quantum computer in the maximally-mixed state is equivalent to obtaining \textit{any} of the possible states with equal probability.
(In other words, a result from a quantum computer in the maximally-mixed state is a `solution' chosen at random from a uniform distribution of all possible solutions.)
Such a `solution' is thus not meaningful, because any accuracy is due to randomness, and not to the performance of the SSO algorithm.

Fig.~\ref{fig:SSO_results_main_text} illustrates that the SSO algorithm offered very little improvement upon such a randomly-determined solution.
For example, in the $N=8$ case, the hardware results offered an improvement of just about 24\%: the result from the maximally-mixed state had an error of $0.71$, the quantum hardware achieved an average error of $0.54$, and so the improvement due to SSO was solely $\frac{0.71-0.54}{0.71}=0.24$.

The fact that SSO's performance on such small problems was so limited illustrates that, although fault-tolerant algorithms like HHL and SSO have significant promise, the noise on contemporary devices is too high for accurately solving even very small problems using these methods.

\subsection{An Algorithm for the Near-Term Era}\label{sec:NISQ_algs}
An alternative to fault-tolerant algorithms are those designed to operate in the NISQ regime, often by leveraging robust classical computing alongside quantum hardware.
Variational Quantum Algorithms (VQAs)~\cite{peruzzo2014variational,cerezo2020variationalreview,bharti2021noisy} encode a task of interest---in our case, solving a linear system---in an optimization problem.
In these algorithms, the classical computer steers the optimization process while the quantum computer computes a cost function, which is being optimized.
The goal is to train a parameterized quantum circuit such that the parameters minimizing the cost function are also those that cause the circuit to compute the solution to the problem of interest.
There are multiple approaches to solving the QLSP in near-term devices~\cite{bravo2020variational,huang2019near,xu2019variational}; we focus on the Variational Linear Solver (VLS) algorithm of Ref.~\cite{bravo2020variational}.
The VLS algorithm trains parameters in a quantum circuit such that, when a cost function is minimized, the solution encoded by the trained circuit is proportional to the solution $\textbf{x}$ of the LSP.

We employed the VLS algorithm to determine pressures at each node in a discretized model of the subsurface.
With VLS, we can currently tackle much more complex problems than we solved with the SSO algorithm.
The problems we considered contained a pitchfork fracture with up to 8192 nodes in the discretization.

\subsubsection{A 6x8 Domain with a Uniform Pitchfork}\label{sec:five_qubit_main_text}
We started with the results of Fig.~\ref{fig:Five_Qubit_Uniform_Perm_Results}, which illustrates that VLS determined the pressures in a 32-node region with a fidelity of greater than 99\%.
Fidelity is a measure of accuracy defined as the inner product between two vectors.
Thus, fidelity is 1---or 100\%---when two vectors have the same direction and proportional magnitude, which equates to a perfect solution in our fracture situation.
(Recall that, since the quantum computer produces a solution vector normalized to 1, the output is proportional to the pressure solution.)
Conversely, fidelity is 0 when two vectors are orthogonal to each other, meaning an entirely inaccurate fracture pressure solution. 
Subfigures (a) and (b) illustrate that the VLS training process---in which we simulated the quantum hardware---generated circuit parameters such that a fidelity of 0.9987 was achieved in the best simulation (highlighted in magenta).
Furthermore, subfigures (c) and (d) illustrate that noise on quantum hardware did not appreciably damage the solution: when running the circuit with the parameters found via optimization, we achieved a fidelity of 0.9911, only 0.0076 away from the fidelity achieved using a noiseless simulator.

Although this is a very small problem when compared to what classical algorithms can accommodate today, this result is significant because it experimentally illustrates that the VLS approach has some resilience to the noise present in NISQ machines.
That in turn suggests why accurate results from quantum computers---even on small problems---are worth exploring.
Quantum computing, both algorithmic and physical implementation, is still in its infancy, so, accurately solving proof-of-concept problems like this one is an important step towards understanding how to make use of quantum computing for fracture systems.

\subsubsection{Larger Domains with Uniform Pitchforks}\label{sec:more_qubits_main_text}
\begin{figure}[ht]
\centerline{\includegraphics[width=\linewidth]{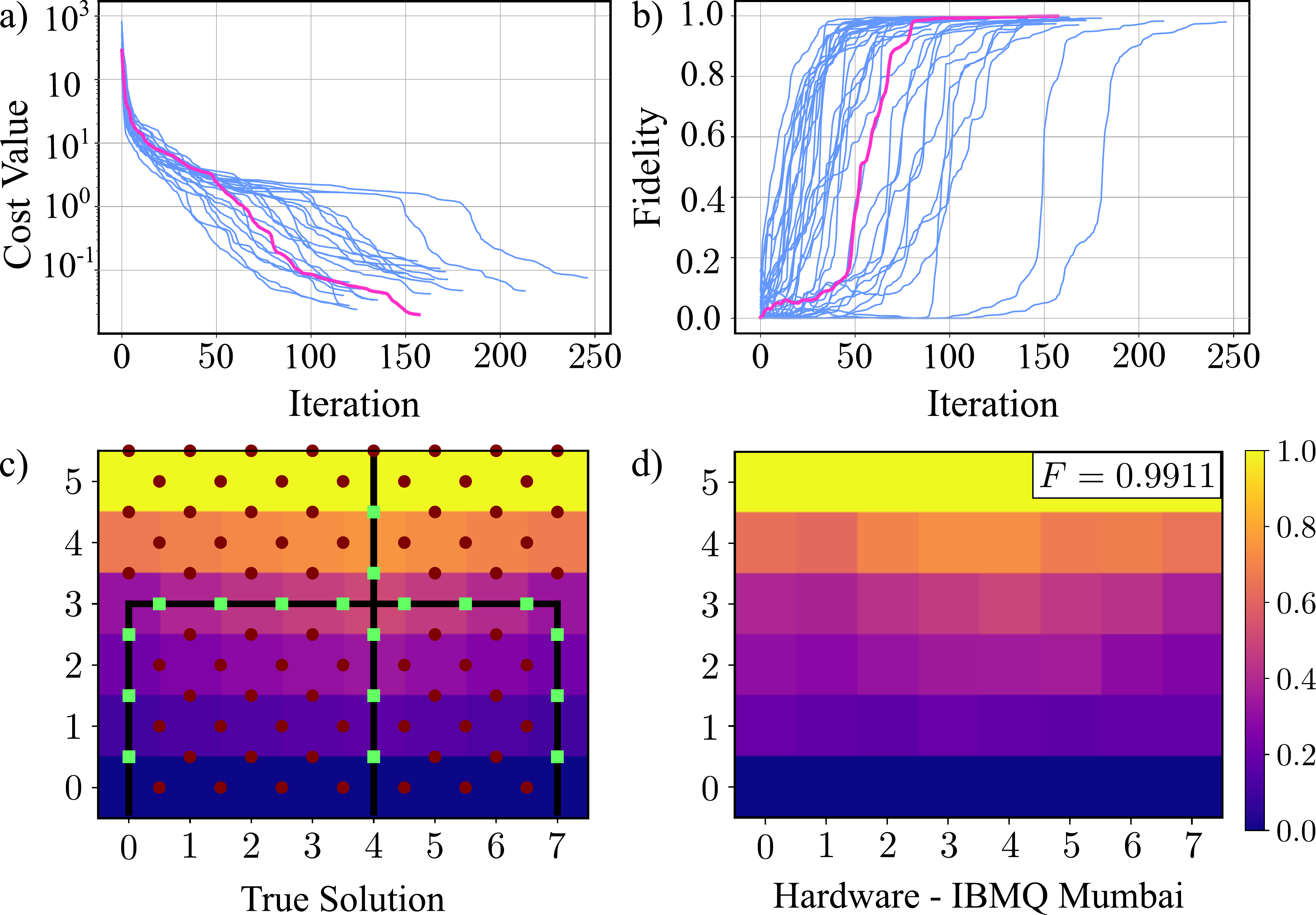}}
\caption{\textbf{Solving a $6 \times 8$ pitchfork fracture problem using a quantum computer.} Subfigures (a) and (b) illustrate the cost and fidelity per iteration for forty sets of randomly-initialized parameters; the result with the highest fidelity is highlighted.  Subfigure (c) illustrates the normalized, known, classically computed solution with overlaid permeabilities.  The inner $4 \times 8$ nodes are the sought-after pressure values because the top and bottom rows have fixed boundary pressures.  The maroon dots illustrate low permeabilities, and the connected green squares illustrate the fracture.  Subfigure (d) is the solution from quantum hardware (specifically, qubits 0, 1, 4, 7, and 10 of the \texttt{ibmq\_mumbai} machine).  This solution has fidelity 0.9911, to four figures.}
\label{fig:Five_Qubit_Uniform_Perm_Results}
\end{figure}

Success with the 32-node problem led us to consider using VLS to solve larger problems.
As predicted, noise affected these solutions more than in the case of Fig.~\ref{fig:Five_Qubit_Uniform_Perm_Results} because increasing region size requires larger circuits---including more qubits and more parameterized quantum gates---to encode the problem.
Nonetheless, we again found that our solutions were quite accurate: the lowest fidelity was 0.8834 for an 8192-node problem. 

\begin{figure}[t]
\centerline{\includegraphics[width=\linewidth]{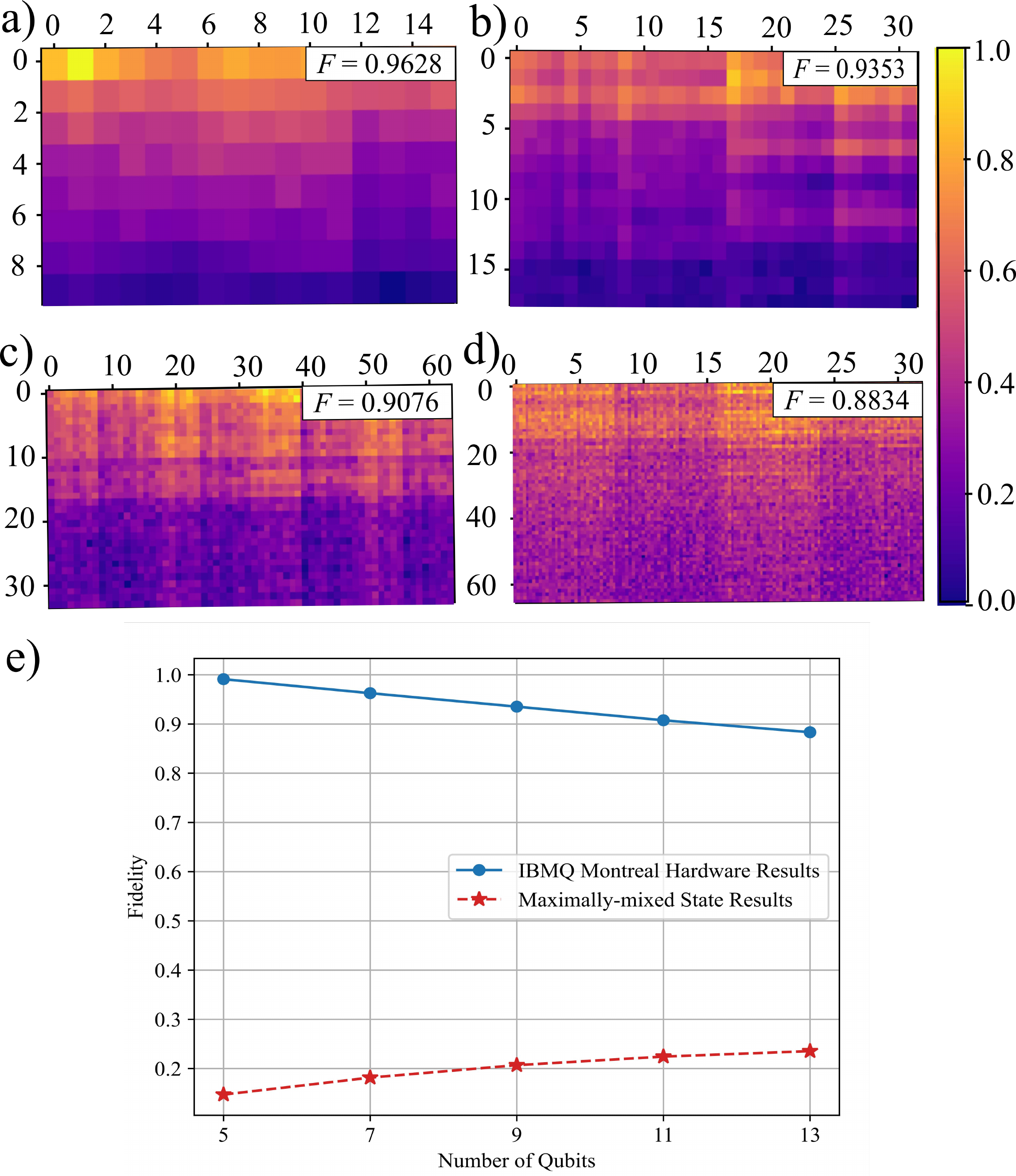}}
\caption{\textbf{Solving a series of larger fracture flow problems.}   Subfigures (a) through (d) are results achieved by running the VLS-trained circuits on the \texttt{ibmq\_montreal} quantum computer for 128-, 512-, 2048-, and 8192- node regions.  (These correspond to 7-, 9-, 11-, and 13-qubit problems, respectively.)  Associated fidelities are 0.9628, 0.9353, 0.9076, and 0.8834.  Subfigure (e) plots fidelities from the same series of problems (in addition to a 5-qubit problem) alongside the fidelity that would have been achieved had the quantum computer's prepared solution degraded to the maximally-mixed state.  The latter illustrates the fidelity from a result comprised solely of random `noise,' thus demonstrating how much better the achieved result on quantum hardware is and experimentally illustrating the noise resilience that is partially proved in Sec.~\ref{sec:noise_resilience}, online.}
\label{fig:More_Qubit_Uniform_Perm_Results}
\end{figure}

Fig.~\ref{fig:More_Qubit_Uniform_Perm_Results} illustrates the details, with subfigure (e) being the most significant result: it indicates that---for all problem sizes considered---we achieved solutions that were significantly more accurate than solutions that had degraded to noise alone.
As in Sec.~\ref{sec:fault_tolerant_algs}, we compared the quality of the solution achieved on quantum hardware to a `solution' that would have been the result of the maximally-mixed state.
And, as in Sec.~\ref{sec:fault_tolerant_algs}, the maximally-mixed state result is a random solution selected from the distribution of all possible solutions.
Unlike with SSO, we found that the quality of VLS's solution was significantly higher than that from the random solution, even for problems that were larger and more complicated than those solved with SSO.
Even the worst fidelity achieved was appreciably above that achieved by a random, noise-only solution: 0.8834 compared to 0.1472.

The performance of VLS on scaled problems was surprising; even as these are relatively small problems, and even as VLS is designed for noisy hardware, we might have seen significantly worse solution quality, as illustrated by Fig.~\ref{fig:More_Qubit_Uniform_Perm_Results}.
This is because, although VLS offloads some computations onto error-proof classical machines, \textit{any} circuits running on contemporary quantum computers are susceptible to noise.
However, quantum algorithms may be less susceptible to noise, if they posses properties that store the relevant information in specific ways, to keep it `protected' from the affects of at least some types of noise.
When we found that the quantum hardware's worst fidelity for scaled problems was appreciably above the associated `noise-only' fidelity, we decided to explore the extent to which the VLS algorithm is noise-resilient~\cite{sharma2019noise}.
In particular, a type of noise known as \textit{depolarizing noise} affects quantum states by making it more likely that they will end up in the maximally-mixed state.
Thus, when we found that VLS solution's fidelity was far above that of the `random' solution, we mathematically established that the VLS algorithm does have at least some resilience to depolarizing noise.
During that process, we also found that that VLS has similarly-limited resilience to what is termed \textit{global dephasing noise}.
Proofs for both of these claims are in Sec.~\ref{sec:noise_resilience}, online.

It is important to clarify that our proofs are solely a first step---albeit an important one---towards completely understanding the noise resilience properties of VLS.
They assume mathematical models of noise that are limited, in the sense that these models do not encompass as many physical situations as can exist.
Specifically, the proofs assume that noise is applied to the quantum state at certain specified locations throughout the circuit, when, in reality, noise could occur at any time during the circuit, including coincidentally with application of a gate operation.
Thus, our proofs are designed to illustrate that VLS does have properties that protect quantum states throughout the algorithm from certain, limited quantum noise patterns.
These proofs, in combination with with the successful empirical results, suggest that further research and empirical evaluation could more completely characterize properties of VLS that offer more expansive noise resilience than the forms which we proved. 

\section{Discussion}\label{discussion}
Quantum computers promise computational improvement for a wide variety of applications, including---as shown in this work---geologic fracture problems.
Although available quantum hardware allows for solving only relatively small problems ($O(10)$ to ($O(10^3)$ nodes), quantum computers are growing and becoming less noisy.
Indeed, there is the potential to begin using fault-tolerant algorithms, such as HHL soon~\cite{smith-goodson2022,chapman2020,kim2021scalable,temme2022}.
Moreover, the development of quantum algorithms better poised to make use of current hardware means quantum computers may be useful for fracture flow problems before the fault-tolerant era arrives.

The algorithms presented in this article suggest that the future of simulating geologic fracture flow might lie with quantum, and our results show that using those algorithms is no longer a solely theoretical consideration: we can now run fracture problems on quantum hardware and obtain relatively accurate results.
Admittedly, these problems are still small, but assuming that the growth and improvement in quantum computers continues as many expect it to---and as it arguably has for the past few decades---we should not be stuck with small problems forever, or even for very long~\cite{smith-goodson2022,chapman2020,vandersypen2001experimental}.
Thus, further experimentation is especially necessary in light of opportunities for further accuracy and scaling, including in less-uniform geologic situations that preliminary investigation suggested were more challenging than their uniform counterparts (See Sec.~\ref{sec:varying_permeability} online.)
Future work should consider tools such as preconditioning~\cite{golden2022quantum}, quantum error mitigation~\cite{li2017efficient,temme2017error,czarnik2020error,o2022near,wang2021can}, `smart encoding,' (see Sec.~\ref{sec:Smart_Encoding} online), problem-specific parametrized quantum circuits~\cite{larocca2022group,pesah2020absence}, and both application- and hardware-specific optimization.
All of these approaches are expected to offer more accuracy and efficiency on larger and more complex problems, thus further establishing the role of quantum computing in the geologic fracture space.

\clearpage

\section{Methods}\label{sec:methods}
\subsection{Adiabatic, Fault-Tolerant Approach}\label{sec:SSO_methods}
The SSO algorithm~\cite{subacsi2019quantum} is inspired by the adiabatic theorem in quantum mechanics, which states that a quantum state will smoothly adapt to changes in its environment if those changes are made sufficiently slowly.
In the context of linear systems, SSO starts with a quantum state that solves a trivial system of equations, and then slowly changes the system into the more complex one whose solution is sought.
SSO changes the system over a discrete sequence of $q$ time steps, and the length of each step is chosen at random from a uniform distribution.
Increasing the number of steps is equivalent to slowing the change of the system, which increases the accuracy of the final solution.

As described in Sec.~\ref{sec:fault_tolerant_algs}, we used the SSO algorithm to solve a linear system specifying two trivial fracture problems.
Both were one-dimensional grids with either $N=4$ or $N=8$ nodes where the left- and right-boundary nodes had fixed pressures.
These conditions---along with the discretized equation in Sec.~\ref{sec:introduction}---specified the $A$ and $b$ for the linear system to be solved.

As Ref.~\cite{subacsi2019quantum} does not provide an explicit quantum circuit implementation of the SSO algorithm, we were limited to creating a unitary matrix representing the net effect of all $q$ steps.
Quantum gates are mathematically represented by unitary matrices, so a single-unitary implementation of SSO is equivalent to a single, large, custom-generated gate aggregating the effect of all $q$ evolutions of the SSO algorithm.
Therefore, for given values of $A$, $b$, and $q$, we generated a unitary matrix via the algorithm described in Ref.~\cite{subacsi2019quantum}.
Because physical implementations of quantum computers cannot run circuits comprised of arbitrary gates, we then broke down that generated matrix into gates that can be executed on existing devices.
To do so, we utilized a variational approach, specifically employing \texttt{Yao.jl}~\cite{luo2019yaojl}, a Julia library for differentiable quantum programming.

The $N=4$ case, which involves only two qubits, was straightforward because any two-qubit unitary matrix can be expressed in terms of a circuit composed of 3 controlled-not and 7 single-qubit gates~\cite{Shende_2004}.
We used the optimization package \texttt{Optim.jl}~\cite{Mogensen2018} to determine the parameter values for the gates to match any given unitary.

The $N=8$ case was more difficult. 
The shortest known universal circuit for three-qubit interactions contains 138 gates~\cite{vatan2004realization}, which is too many for consistently-accurate performance on existing hardware. 
We therefore employed a machine learning approach across circuits of increasing gate count until we were able to find a circuit that matched the unitary to a high degree.
We were regularly able to find circuits with 50 gates (approximately 30 single-qubit gates and 20 controlled-not gates) that achieved at least $99.67\%$ fidelity.
Circuits with fewer gates resulted in poor performance.

Once we had obtained circuits that implemented SSO for our fracture systems, we ran them on IBMQ's suite of quantum computers. 
Specifically, we used the \texttt{ibmq\_qasm\_simulator} to simulate performance on a hardware-noise-free quantum device, and then we compared with performance on the quantum computer \texttt{ibmq\_rome}.

Quantum computers---and therefore the algorithms that work thereon---are inherently probabilistic.
So, most quantum algorithms require running a circuit many times and `measuring' the resulting state each time to establish a probability distribution of states.
The probability that each state occurs provides the vector of solutions for the problem that the quantum algorithm sought to solve.
Each run/measurement combination is termed a ``shot,'' and we ran each SSO circuit with 8192 shots on both the simulator and hardware.
Using the results, we could then infer the observed value of our sought-pressure solution, $\textbf{x}$.

Due to the stochastic nature of the algorithm (i.e., randomly-chosen time lengths, $q$), we averaged performance over 75 instances (i.e., distinct time-evolution sequences generated for fixed values of $A$, \textbf{$b$}, and $q$).  Fig.~\ref{fig:SSO_results_main_text} depicts the results, which are also described in Sec.~\ref{sec:fault_tolerant_algs}.

Finally, we computed the error that would have occurred had the `solution' in the quantum computer degraded to noise alone.
We did this by considering the mathematical representation of the maximally-mixed state, which is a state that contains solely noise.
The maximally-mixed state can be represented as a density matrix, $\rho$:
\begin{equation}\label{eq:maximally-mixed-density-operator}
    \rho = \frac{1}{2^n}I_{n\times n}, 
\end{equation}
where $n$ is the number of qubits~\cite{nielsen2000quantum}.
(Note: Throughout Secs.~\ref{sec:methods} and ~\ref{sec:supplementary_info}, we use density operators and Dirac notation, both of which are standard notation for mathematically representing quantum circuits.  For a thorough introduction to density operators, please see Ref.~\cite{nielsen2000quantum}.  For an introduction to Dirac notation, please see Ref.~\cite{diracIntro}.)
When the maximally-mixed state is measured (which can be mathematically represented as projecting the state onto a specified basis), we obtain a result that is equivalent to selecting a pressure at each node randomly from a uniform distribution of all possibilities.

We can illustrate this by considering the probability of measuring a certain 2-qubit state such as that used for the $N=4$ problems.
The probability of measuring a given state, $m$ from a quantum circuit represented by density operator, $\rho$ is given by,
\begin{equation}\label{eq:measuring-maximally-mixed-state}
    p(m) = Tr(M_m^\dagger M_m \rho),
\end{equation}
where $M_m$ is the measurement operator for a given basis.
(For an introduction to quantum measurement, please see Ref.~\cite{nielsen2000quantum}.)
We seek the probabilities of measuring $00$, $01$, $10$, and $11$; in our fracture flow problem, each of these probabilities corresponds to the pressure in one of the nodes.
For the computational basis, which we used for the results in this paper, the measurement operators, $M_m$ for each of the above possible solutions are $M_{00}=\ket{00}\bra{00}$, $M_{01}=\ket{01}\bra{01}$, $M_{10}=\ket{10}\bra{10}$, and $M_{11}=\ket{11}\bra{11}$.
So, when the state of the circuit, $\rho$, is equivalent to the maximally-mixed state for 2 qubits (i.e., a $4 \times 4$ identity matrix with a coefficient of $\frac{1}{4}$), we have that the probabilities for each possibility are given by $\frac{1}{4}.$
We thus see that---for a circuit in the maximally-mixed state---the probabilities of all possible states have been reduced to the same value, meaning the `solution' of the maximally-mixed state contains no meaningful information.
Any resemblance to our desired solution, $\textbf{x}$, is the result of random chance and not the performance of an algorithm.
So, to benchmark SSO against the results of random chance, we computed the error that would have occurred had the quantum computer's returned `solution' been one of random chance alone given degradation to the maximally-mixed state.

\subsection{Variational Linear Solver Approach}\label{sec:VLS_methods}
\subsubsection{Introduction to VLS}
\begin{figure}[t]
\centerline{\includegraphics[width=\columnwidth]{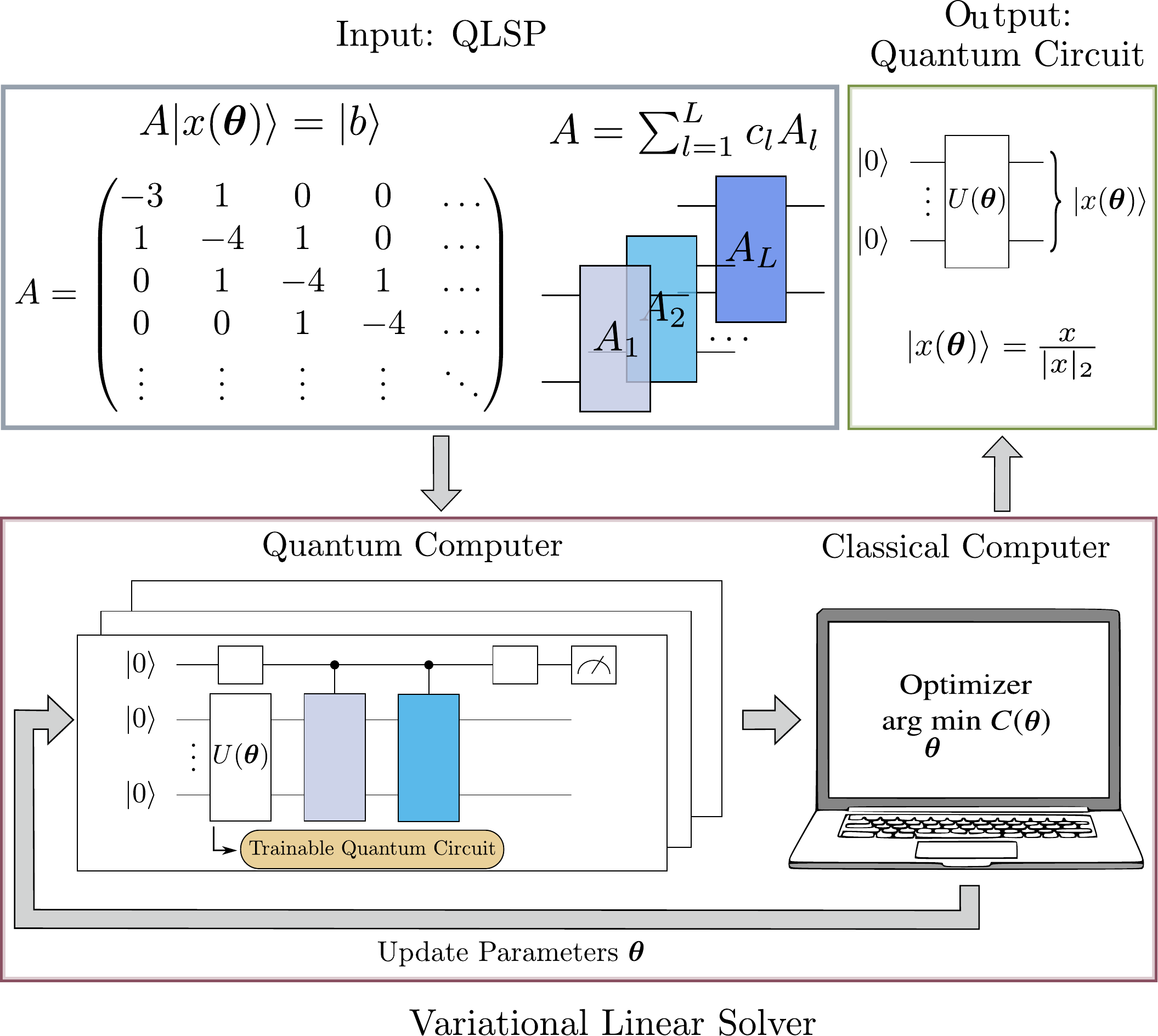}}
\caption{\textbf{Variational Linear Solver algorithm.} As described in Fig. \ref{fig:VLS_for_Geosciences_Workflow}, the algorithm accepts as input an $A$ and $\textbf{b}$ specifying a LSP and minimizes a cost function to create a parameterized quantum circuit that solves the LSP.  Specifically, the cost function optimization takes advantage of classical optimization techniques, while the cost function evaluation occurs via either quantum hardware or a classical simulator of such hardware.} 
\label{fig:VLS_Algorithm}
\end{figure}

As is schematically shown in Fig.~\ref{fig:VLS_Algorithm}, the VLS algorithm takes a description of the QLSP (i.e., $A$ and $\boldsymbol{b}$) as input.

\begin{figure}[t]
\centerline{\includegraphics[width=.8\columnwidth]{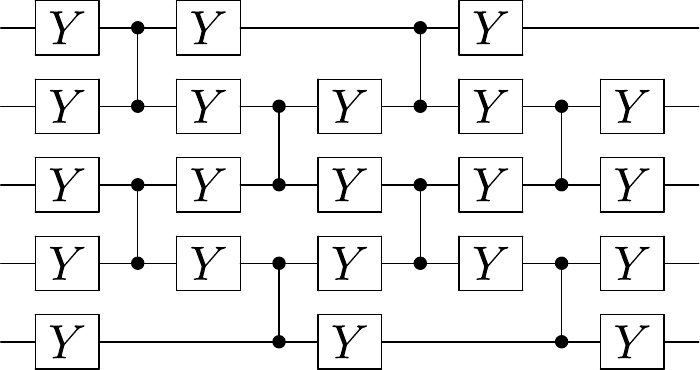}}
\caption{\textbf{Five-qubit, two-layer ansatz.}  The pattern for this ansatz is a ‘preliminary’ layer of $R_y$ gates followed by layers with two sets of controlled-$Z$ and $R_y$ gates each.  Each $R_y$ gate’s angle is one of the tuned parameters.}
\label{fig:Two_Layer_Five_Qubit_Ansatz}
\end{figure}

To solve the QLSP, the VLS algorithm trains the parameters $\boldsymbol{\thv}$ in a quantum circuit, $U(\boldsymbol{\thv})$.
Fig.~\ref{fig:Two_Layer_Five_Qubit_Ansatz} illustrates the ansatz structure of the quantum circuit that we sought to train with the VLS algorithm.
The circuit contains (unparameterized) controlled-$Z$ gates and parameterized single qubit rotations about the $y$-axis. Thus, the parameter $\theta_i\in\boldsymbol{\thv}$ corresponds to a trainable rotation angle in the $i$-th rotation such that $\theta_i\in[0,2\pi]$~\cite{bravo2020variational}.
We chose this ansatz because it had been used successfully with the VLS algorithm in previous work~\cite{bravo2020variational} and because it is `hardware-efficient,' meaning it uses gates whose structures offer the lowest error-rates available on current NISQ devices.

The  $U(\boldsymbol{\thv})$ circuit prepares a trial solution $\ket{x(\boldsymbol{\thv})}=U(\boldsymbol{\thv})\ket{\vec{0}}$, where $\ket{\vec{0}}$ is a state in which all qubits are initialized to the easy-to-prepare initial state, $\ket{0}$.
To calculate the quality of the resulting quantum state $\ket{x(\boldsymbol{\thv})}$ as a solution to the QLSP, VLS minimizes a cost function $C(\boldsymbol{\thv})$ that quantifies how much each component of $A\ket{x(\boldsymbol{\thv})}$ is orthogonal to $\ket{b}$.
It can be verified that the cost function
\begin{equation}\label{eq:Cost_Func}
    C(\boldsymbol{\thv})=\bra{x(\boldsymbol{\thv})}H\ket{x(\boldsymbol{\thv})}\,, 
\end{equation}
with
\begin{equation}\label{eq:Hamiltonian}
    H=A\ad (\id-\ket{b}\bra{b})A\,,
\end{equation}
is minimized if and only if $\ket{x(\boldsymbol{\thv)}}$ is proportional to the solution $\textbf{x}$ of the LSP~\cite{bravo2020variational}. Note that, here, VLS maps the QLSP into a problem of finding the ground-state of the Hamiltonian given in eq.~\eqref{eq:Hamiltonian}.

Once the minimization task, $\argmin_{\boldsymbol{\thv}} C(\boldsymbol{\thv})$,
is solved, the VLS output is a parameterized quantum circuit that prepares a quantum state $\ket{x(\boldsymbol{\theta_{final}})}$ that approximates $\boldsymbol{x}/|\boldsymbol{x}|_2$
As mentioned in Sec.~\ref{sec:SSO_methods}, obtaining these values requires performing measurements on (i.e., collecting shots from) the state $\ket{x(\boldsymbol{\theta_{final}})}$ to obtain a vector of estimated probabilities that represents a solution to the LSP.
Specifically, expressing the solution as 
\begin{equation}\label{eq:Computational_basis}
    \ket{x(\boldsymbol{\theta_{final}})}=\sum_i\frac{x_i}{\sqrt{\sum_i |x_i|^2}}\ket{\vec{z}_i}\,,
\end{equation}
where each $x_i$ is an element of $\ket{x}$ and  $\{\ket{\vec{z_i}}\}_{i=1}^{2^n}$ are the elements of the  computational basis such that $\vec{z}_i\in\{0,1\}^{\otimes n}$, then the values $|x_i|^2$ correspond to the pressures at the nodes in the discretized surface.
With sufficiently accurate $\boldsymbol{\theta_{final}}$ parameters and enough samples, the vector of estimated $x_i$'s can be brought within a tolerated error of the elements in the desired $\textbf{x}$.

To assess the quality of a given solution obtained by VLS, there are two approaches that depend upon whether the solution to the problem is known.
If the desired vector, $\boldsymbol{x}$ is unknown, then the value of the cost function, which takes as input a state $\ket{x}$ generated by a given set of parameters, can be used to evaluate the quality of the final parameters sent as input.
In this case, the goal is simply to make the cost function evaluate to as small (i.e., close to zero) a value as possible.

Conversely, if the desired vector, $\boldsymbol{x}$ is known---as it was in the experiments we designed---then the quality of parameters can be assessed using a unitless quantity termed \textit{quantum fidelity}.
Quantum fidelity is defined as the inner product between two vectors,
\begin{equation}\label{eq:Fidelity}
    F(\ket{x},\ket{x_{\text{true}}})=|\braket{x}{x_{\text{true}}}|^2\,,
\end{equation}
where $F(\ket{x},\ket{x_{\text{true}}})=1$ if and only if $\ket{x}$ is equal to $\ket{x_{\text{true}}}$ (up to a global unmeasurable phase), and $F(\ket{x},\ket{x_{\text{true}}})=0$ if the two states are orthogonal.
Because our goal was to assess the performance of VLS, we solved problems for which we had classically-computed true solutions, meaning we computed the quantum fidelity between the solution obtained by VLS, $\ket{x}$, and a state representing the normalized, true solution $\ket{x_{\text{true}}}$.

\begin{figure}[t]
\centerline{\includegraphics[width=.6\columnwidth]{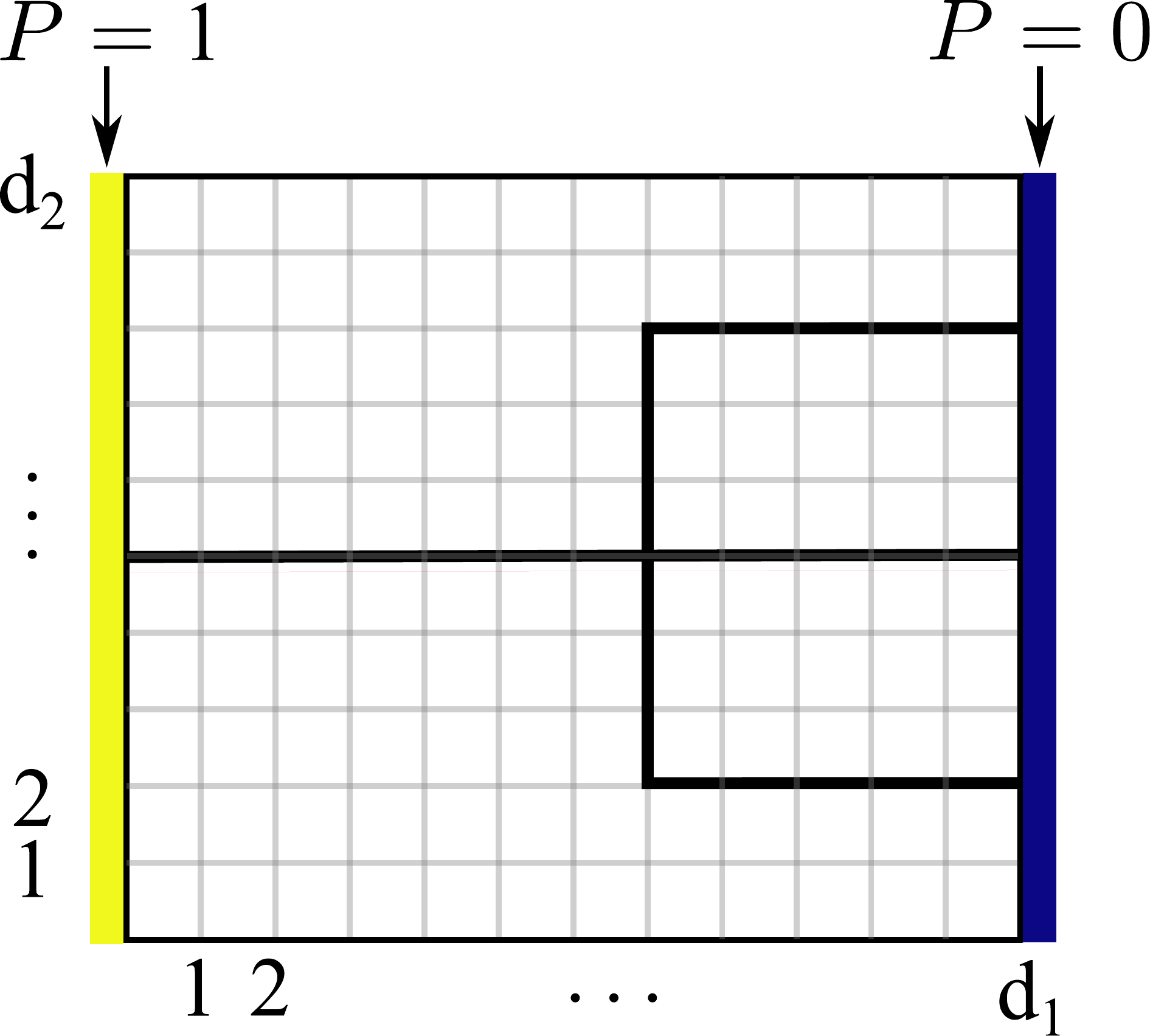}}
\caption{\textbf{Pitchfork and surface discretization.}  A cartoon of the subsurface flow situation, in which a region is discretized into $d_1 \times d_2$ nodes.  Pressure boundary conditions of 1 and 0 are imposed on the left- and right-hand sides of the region, respectively, and a pitchfork fracture runs through the middle of the region.  Each node has a pressure that we seek as our solution.}
\label{fig:Pitchfork_Situation_Cartoon}
\end{figure}

\subsubsection{Uniform Permeability with 6x8 Region}
We first considered a uniformly-permeable pitchfork embedded in a $6\times 8$ grid.
As shown in Fig.~\ref{fig:Pitchfork_Situation_Cartoon}, the subsurface for a given problem needs to be discretized into a grid of size $d_1\times d_2$.
For the problem to fit into an $n$-qubit quantum state, the size of the grid needs to be a power of $2$, meaning one needs to choose a grid size such that $d_1 d_2=2^n$.
In this case, $d_1\times d_2$ did not equal $6\times 8$, but instead $4 \times 8$.
Because we imposed Dirichlet boundary conditions, the pressures depicted on the left- and right-most edges of Fig.~\ref{fig:Pitchfork_Situation_Cartoon} were fixed at one and zero, respectively.
Thus, the solution of interest contained the pressures of only each inner $4 \times 8$ grid node, giving a linear system with $A$ of dimension $32\times 32$ and with $\boldsymbol{x}$ and $\boldsymbol{b}$ of dimension $32 \times 1$.
As $2^5=32$, this pitchfork problem was solved using five qubits.

While our approach allowed for the branches of the pitchfork fracture to have varying permeabilities (see Sec.~\ref{sec:varying_permeability} online), we first considered a uniformly-permeable pitchfork that had a permeability ten times greater than that of the surrounding surface.

During the training phase of VLS, we ran 40 instances of the algorithm, where the trainable parameters $\boldsymbol{\thv}$ were randomly initialized at each instance.
Each of these instances included multiple iterations, where each iteration corresponds to the classical optimizer taking as input the value of the cost function, and producing an updated set of parameters (see Fig.~\ref{fig:VLS_Algorithm}).
We used Scipy's optimize package (specifically, \texttt{minimize}) with the conjugate-gradient method~\cite{scipy-optimize-cg} on a five-layer ansatz of the form described in Ref.~\cite{bravo2020variational} and illustrated in Fig.~\ref{fig:Two_Layer_Five_Qubit_Ansatz}.
The cost function was evaluated using a classical simulator with shot noise (meaning we used a limited number of shots, and, specifically, $10^8$) but without simulated hardware noise.
The cost function was evaluated as described in Ref.~\cite{cerezo2020variationalreview}.
In Fig.~\ref{fig:Five_Qubit_Uniform_Perm_Results}, subfigures (a) and (b), the cost and fidelity per iteration are plotted for each of the forty instances.
The fact that the fidelities per iteration in subfigure (b) converge to one indicates that VLS was able to find the solution of the QLSP.
The instance highlighted in purple obtained the highest-fidelity results; after 150 iterations, it achieved a cost function value less than $10^{-1}$, and a fidelity of $0.9987$.

We then ran the quantum circuit with the highest-fidelity parameters on quantum hardware and, specifically, qubits 0, 1, 4, 7, and 10 of the \texttt{ibmq\_mumbai} machine.
These qubits were selected both for their connectivity and relatively low error rates.
First, connectivity: it is sensible to select topologically-connected qubits to take advantage of the hardware-efficient structure of the circuit.
Otherwise the final circuit would involve additional gates, meaning higher-than-necessary amounts of noise and higher-than-necessary possibility for error.
As our goal was to obtain the highest fidelity possible despite the imperfections of existing hardware, we chose qubits that were topologically-connected.
Second, error-rate: amongst the connected sets of five qubits available, we chose 0, 1, 4, 7, and 10 because that group avoided inclusion of qubits with high readout assignment errors, high single gate errors, and high controlled-not (i.e., two-qubit) gate errors.
We obtained error information via IBMQ's hardware dashboard; this information changes in real time due to continual calibration of the machines.

We used approximately $10^5$ shots; because the \texttt{ibmq\_mumbai} machine has a single-circuit shot maximum of 8192, we ran the circuit 12 times with 8192 shots each time for a total of 98,304 shots.
Finally, we permitted Qiskit to perform the maximum number of optimizations allowable by setting the optimization\_level flag to three. (Qiskit provides varying automated levels of optimization on a scale of 0---no optimization---to 3---as much optimization as possible~\cite{qiskit-optimization}.)

In Fig.~\ref{fig:Five_Qubit_Uniform_Perm_Results}, subfigures (c) and (d) illustrate the performance of VLS with the five-layer ansatz and the parameters found in the highest-fidelity instance highlighted in subfigures (a) and (b).
Subfigure (c) illustrates the pressure grid corresponding to the normalized, known, true solution of the LSP, as well as the discretized pitchfork fracture as points on the edges of the grid.
Subfigure (d) depicts the pressure solution obtained from the quantum hardware.  
As previously mentioned, algorithms solving QLSPs prepare a solution that is \textit{proportional} to the solution $\boldsymbol{x}$ of the LSP, which preserves the relationship between the elements of vector solution $\boldsymbol{x}$.
We chose to plot the normalized true solution to more clearly visualize that the relationship between solution elements was indeed preserved in the quantum computer's solution.
As is described in Sec.~\ref{sec:NISQ_algs}, subfigure (d) indicates that hardware noise did not significantly disrupt the circuit's ability to compute an accurate solution; the quantum hardware generated a solution with fidelity 0.9911.

\subsubsection{Uniform Permeability with Larger Regions}
We next considered VLS's scalability on pitchfork fracture problems.
For quantum states $\ket{x(\boldsymbol{\thv})}$ of larger dimensions, we can determine the suitability of the solution by minimizing the cost function $C(\boldsymbol{\thv})$ in eq.~\eqref{eq:Cost_Func} as before.
However, this direct approach is computationally challenging for circuits with increasingly many qubits.
Thus, to simplify the training computation, the minimization of $C(\boldsymbol{\thv})$ may be replaced by globally minimizing a new cost function,
\begin{equation}\label{eq:Cost_Func_Large}
    \tilde{C}(\boldsymbol{\thv})= 1 - \left|\langle x_{\text{true}} \ket{x(\boldsymbol{\thv})} \right|^2\,, 
\end{equation}
where $\ket{x_{\text{true}}}$ is the solution vector of true pressures.
Both equations \eqref{eq:Cost_Func} and \eqref{eq:Cost_Func_Large} achieve minima when $\ket{x(\boldsymbol{\thv})} \approx \ket{x_{\text{true}}}$.
Because our goal was to evaluate the performance of VLS, we used only problems for which we could in fact obtain a classical solution to compare against, and this meant that we could classically obtain $\ket{x_{\text{true}}}$ for all of the problems in this article.
Thus, we could apply the less-computationally-intense cost function formula above during the training phase.

During the VLS training phase, we once again began with randomly-initialized parameters, and each iteration of the training corresponded to the classical optimizer taking the current value of the cost function to produce an updated set of parameters.
Again, we used Scipy's \texttt{minimize} with the conjugate-gradient method, and, again, we trained with shot noise ($15*10^{13}$ shots), but no hardware noise.
It is worth noting that the number of shots required to train the circuits such that $\tilde{C}(\boldsymbol{\thv}) < 10^{-3}$ (roughly corresponding to fidelities near or above 0.9) increased dramatically for larger problems, which is in part due to the significantly larger circuits that had to be trained.
Not only were there more qubits (7, 9, 11, or 13), but because the problems were larger, the circuits also contained more parameterized gates.
Experimentation illustrated that a number of ansatz layers greater than or equal to the number of qubits trained circuits well, so we chose the number of ansatz layers to equal the number of qubits.

\begin{figure}[t]
\centerline{\includegraphics[width=.8\columnwidth]{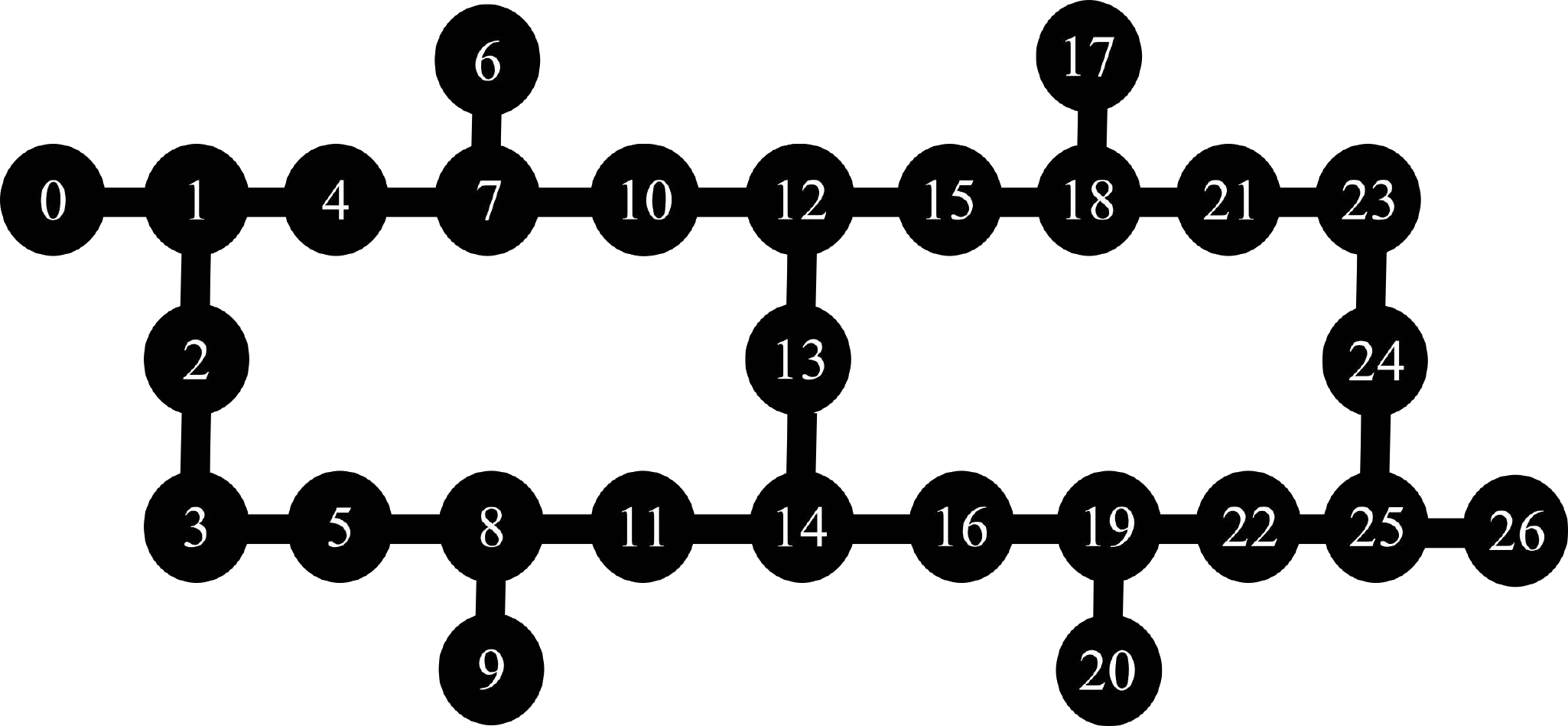}}
\caption{\textbf{The qubit connectivity of IBM's Montreal quantum computer.}}
\label{fig:IBMQ_Montreal_Layout}
\end{figure}

We then ran the quantum circuit with the highest-fidelity parameters for each of the differently-sized problems on quantum hardware.
We again used approximately $10^5$ shots, this time rounding up to 13 runs of 8192 shots each.
The qubit selection procedure was more complex because, when selecting five qubits for the smaller problem, it was straightforward to choose a group that avoided the worst-performing qubits.
Moving up to even the seven-qubit problem made the selection task more difficult because it was no longer obvious which sets would best reduce error; for example, would it be preferable to include one qubit with very poor performance, or two qubits with better---but still bad---performance?
We opted to address this qubit-selection challenge by trying many qubit combinations for each of the $n= 7, 9, 11,$ or $13$ qubit problems.
Specifically, we used each possible set of $n$ qubits in which the qubits were adjacent to each other and did not `double-count' any given qubit.  We undertook this procedure for each size of problem on the \texttt{ibmq\_montreal} machine, and to help clarify the process, Fig.~\ref{fig:IBMQ_Montreal_Layout} illustrates the connectivity of \texttt{ibmq\_montreal}.
Consider the seven-qubit problem: qubits 6, 7, 10, 12, 13, 14, and 16 were a possible qubit selection, but qubits 4, 7, 6, 10, 13, 14, and 16 were not, because the latter would require `double-counting' qubit 7 while determining qubit adjacency.

Fig.~\ref{fig:More_Qubit_Uniform_Perm_Results} summarizes our results, illustrating the highest-fidelity pressure solution obtained for the 7, 9, 11, and 13-qubit problems.
In particular, the highest-fidelity results occurred for the 7-qubit problem with qubits 3, 5, 8, 11, 12, 13, and 14; for the 9-qubit problem with qubits 0, 1, 2, 3, 5, 8, 11, 13, and 14; for the 11-qubit problem with qubits 5, 8, 11, 14, 16, 19, 21, 22, 23, 24, and 25; and for the 13-qubit problem with qubits 8, 9, 11, 14, 15, 16, 18, 19, 21, 22, 23, 24, and 25.
As described in Sec.~\ref{sec:NISQ_algs}, subfigure (e) presents results regarding the quality of the solution when compared to a solution containing solely noise.

\section{Data Availability}\label{data_availability}
The data for generating the figures (excepting those illustrating cartoons/concepts) is available at https://github.com/JessieMHenderson/quantum-geologic-fracture-networks.git.
Instructions for generating figures from the data can be obtained from the corresponding author upon reasonable request.

\section{Code Availability}\label{code_availability}
The code used to generate the results is available at https://github.com/JessieMHenderson/quantum-geologic-fracture-networks.git.
Instructions for utilizing the code (beyond those contained within the documentation) can be obtained from the corresponding author upon reasonable request.

\bibliography{References}\label{references}

\section{Acknowledgements}\label{sec:acknowledgements}
JMH, JKG, DO and HSV gratefully acknowledge support from the Department of Energy, Office of Science, Office of Basic Energy Sciences, Geoscience Research program under Award Number (LANLE3W1). MP was
supported by the U.S. DOE through a quantum computing program sponsored by the Los Alamos National
Laboratory (LANL) Information Science \& Technology
Institute. MC acknowledge initial support from the Center for Nonlinear Studies at Los Alamos National
Laboratory (LANL). MC was also supported through LANL's ASC Beyond Moore’s Law project.
DO, JKG and HSV gratefully acknowledge support from Los Alamos National Laboratory's Laboratory Directed Research \& Development project 20220077ER.

\section{Author Information \& Ethics Statements}\label{sec:author_information_and_ethics}
\subsection{Authors and Affiliations}
\begin{itemize}
  \item Jessie M. Henderson: Los Alamos National Laboratory and Southern Methodist University
  \item Marianna Podzorova: Los Alamos National Laboratory, Joint Center for Quantum Information \& Computer Science, and University of Maryland
  \item Marco Cerezo: Los Alamos National Laboratory
  \item John K. Golden: Los Alamos National Laboratory
  \item Leonard Gleyzer: Los Alamos National Laboratory and Brown University
  \item Hari S. Viswanathan: Los Alamos National Laboratory
  \item Daniel O'Malley: Los Alamos National Laboratory
\end{itemize}

\subsection{Author Contributions}\label{sec:author_contributions}
HSV, DO, and MC designed the project.
DO wrote the code for generating the problems used in Secs.~\ref{sec:fault_tolerant_algs} and~\ref{sec:NISQ_algs}.
LG and JKG wrote the code and collected the data for the SSO algorithm in Sec.~\ref{sec:fault_tolerant_algs}.
MP and JMH wrote the code and collected the data for the VLS algorithm in Sec.~\ref{sec:NISQ_algs} under the direction of MC, who contributed substantially to this code and to analysis of the resulting data.
MC developed the mathematics of the project, including the proofs of noise resilience; JMH, MC, and MP wrote them as presented in Sec.~\ref{sec:noise_resilience}.
MC and JMH created the figures.
JMH wrote the first draft of the manuscript, with significant guidance from HSV, DO, MC, and JKG.
All authors discussed, commented upon, and revised the final manuscript.

\subsection{Materials and Correspondence}\label{sec:corresponding_author}
Please contact Jessie Henderson (jessieh@lanl.gov) with questions or comments.

\subsection{Competing Interests}\label{sec:competing_interests}
The authors declare no competing interests.

\clearpage

\section{Supplementary Information}\label{sec:supplementary_info}
\subsection{Theoretical Results on Noise Resilience}\label{sec:noise_resilience}
\subsubsection{Brief Introduction to Quantum Noise}
As has been emphasized, current quantum hardware is noisy, meaning it cannot yet sufficiently avoid information loss caused by uncontrolled interactions with the environment or other qubits.
This means that the fragile entities we use as qubits (whether those be ions, photons, or microwave-controlled superconducting chips) are changing in such a way that the information we tried to store is damaged~\cite{tqi2020}.
Mathematically, we can represent noise as we can any other operation on a qubit---the only difference is that such noise operations are uncontrolled and determined by qubit-qubit or qubit-environment interaction.

By leveraging an algorithm like VLS, in which a portion of the computation is performed on classical computers, we reduce the amount of quantum noise: because less time is spent on a quantum computer, there is necessarily less opportunity for noise exposure.
However, any circuit that is run on a quantum computer is susceptible to noise during that process.
So, as long as there is any `quantum portion' of a hybrid algorithm running on a NISQ machine, exposure to noise is guaranteed.
We can mitigate the effects of such inevitable noise by modifying how quantum algorithms store the information most relevant to the ultimate solution.
When an algorithm avoids damage from a type of quantum noise through such design features, the algorithm can be said to be \textit{resilient to} that sort of noise.

Before going further, it is worth emphasizing that \textit{resilience} to quantum noise should not be confused with \textit{resistance} to quantum noise, where the latter connotes immunity rather than simply some measure of defense.
Demonstrating mathematical resilience to a type of quantum noise does not mean that that sort of noise will have \textit{no} effect on a quantum state because mathematical models always depend upon assumptions.
And, as described in Sec.~\ref{sec:NISQ_algs}, our proofs are only a first step towards completely characterizing the noise resilience of VLS that has been empirically demonstrated.
Our assumptions require that the noise occur at certain times throughout the circuit operation, and while those assumptions do allow us to prove VLS has resilience to noise in those cases, we make no claims about even the same types of noise occurring in other locations throughout the circuit.

We do believe, however, that there is more to be shown regarding the noise resilience of VLS.
We empirically demonstrated that VLS has far more noise resilience than could be explained by the limited proofs herein.
Therefore, future work should further explore the algorithm's noise resilience both mathematically and experimentally: while formal mathematics illustrates that a certain set of experiments was not simply a coincidence, empirical successes show that the assumptions on which the mathematical models were based have sufficient accuracy to derive practically-useful conclusions.

Before providing the theorems and proofs, we provide a brief introduction to the mathematical representation of quantum noise.
(For a thorough introduction to quantum noise, please see Ref.~\cite{nielsen2000quantum}.)
As mentioned previously, quantum states can be represented mathematically using matrices known as density-operators~\cite{nielsen2000quantum}.
These provide a more general---albeit sometimes less intuitive---notation than the statevector notation used most often in this article~\cite{nielsen2000quantum}.
Density-operators can clearly express the state of a quantum circuit even when we lack complete information about what operations have occurred.
Specifically, if some operation will occur with probability $p$ and another with probability $1-p$, density operators can encapsulate these possibilities on the level of circuits and not individual qubits.

A density operator is defined as,
\begin{equation}\label{eq:Density_Operator_Definition}
    \rho = \sum_i p_i\ket{\psi_i}\bra{\psi_i},
\end{equation}
where each $p_i$ is the probability that the quantum system is in one of $i$ states from a set $\{\ket{\psi_i\}}$~\cite{nielsen2000quantum}.
We can then represent the evolution of quantum states using operator-sum notation, which breaks a quantum operation (e.g., application of quantum noise that might---or might not---occur) into a summation that we can use to `build up' descriptions of system-wide operations.
Specifically, we can write,
\begin{equation}\label{eq:Density_Operator_Definition_E_k}
    \varepsilon(\rho) = \sum_k E_k\rho E_k^\dagger,
\end{equation}
where $\varepsilon(\rho)$ is the system-wide quantum operation applying to system state $\rho$, and the $E_k$'s represent matrices applying pieces of that quantum operation.

Quantum noise can thus be represented as a set of $E_k$'s, meaning we can build a quantum operator $\varepsilon$ to represent application of quantum noise.
The use of density operators and operator-sum notation allows for representing quantum states in which there is uncertainty about whether or not noise will occur.

There are at least four types of quantum noise, all of which have different physical---and thus mathematical---representations.
One general noise phenomenon is known as \textit{decoherence}, which refers to loss of information when a qubit evolves towards a state that nature `prefers' (e.g., a state where the qubit's temperature matches that of the environment)~\cite{nielsen2000quantum}.
Two types of decoherence are dephasing noise and global depolarising noise.
We focus on these here because they are the two types to which the VLS algorithm exhibits partial resilience in certain situations.

\subsubsection{Dephasing Noise}
Dephasing noise acting on a single qubit is mathematically represented in operator-sum notation with,
\begin{equation}\label{eq:Dephasing_Noise_E0}
    E_0 = \sqrt{p}\begin{pmatrix}
		  1 & 0 \\
		  0 & 1
		\end{pmatrix} = \sqrt{p}I
\end{equation}
and
\begin{equation}\label{eq:Dephasing_Noise_E1}
	 E_1 = \sqrt{1-p}\begin{pmatrix}
		  1 & 0 \\
		  0 & -1
		\end{pmatrix}  = \sqrt{1-p}Z,
\end{equation}
where $p$ is the probability that the noise does not affect state $\rho$, $I$ is the identity gate, and $Z$ is the Pauli-Z gate~\cite{nielsen2000quantum}.

\begin{theorem}\label{th:Dephasing_Noise_Resilience}
If dephasing noise occurs at the end of the circuit, the VLS algorithm is resilient to that noise, because dephasing noise affects the off-diagonal elements of a quantum state, $\rho$, while the VLS algorithm stores computed information in the diagonals of $\rho$.
\end{theorem}

\begin{specialproof}
\begin{corollary}\label{th:Dephasing_Noise_Single_Qubit}
The VLS algorithm is resilient to single-qubit dephasing noise applied at the end of a quantum circuit.
\end{corollary}

\begin{specialproof}
Considering the definition of dephasing noise as provided in eqs.~\eqref{eq:Dephasing_Noise_E0} and ~\eqref{eq:Dephasing_Noise_E1}, the density matrix of the dephasing noise operation, $\tilde{\rho}$ on a single qubit of a quantum state $\rho$ is
\begin{equation}\label{eq:Dephasing_Noise_Density_Matrix}
    \tilde{\rho} = \varepsilon_d(\rho) = pI\rho I^\dagger + (1-p)Z\rho\Z^\dagger,
\end{equation}.
Here, $p$ is the probability that dephasing noise does not disrupt $\rho$, while $(1-p)$ is the probability that it does.
Simplifying gives,
\begin{equation}\label{eq:Dephasing_Noise_Density_Matrix_Simplified}
    \tilde{\rho} = p\rho + (1-p)Z\rho Z.
\end{equation}

Thus, we see that $\varepsilon_d$ will meaningfully change only the elements on the diagonal of $\rho$, given the second term in eq.~\eqref{eq:Dephasing_Noise_Density_Matrix_Simplified}.
Because $Z$ is diagonal with $\pm 1$, it flips the signs of rows corresponding to the location of its $-1$ when it is left-multiplied with another matrix and flips the signs of columns corresponding to the same when it is right-multiplied with another matrix.
Eq.~\eqref{eq:Dephasing_Noise_Density_Matrix_Simplified} applies $Z$ from both sides, meaning those sign flips will cancel each other out, leaving the diagonal elements of $\rho$ unchanged.

Recall that the goal of the VLS algorithm is to estimate $|c_i|^2$ in $\ket{x(\thv)}=\sum_i c_i \ket{\vec{z_i}}$.
Because $\vec{z_i}$ is the set of canonical basis vectors, we see that the terms involving $|c_i|$ are on the diagonal of the specified matrix.
Thus, the most valuable information in our quantum state lies on the diagonal of the density operator representing that state:
\begin{equation}\label{eq:Density_Operator_VLS)}
	\rho = \ket{x(\thv)}\bra{x(\thv)}=\sum_i |c_i|^2 \ket{\vec{z_i}}\bra{\vec{z_i}} + \sum_{i\ne j} c_i c_j^* \ket{\vec{z_i}}\bra{\vec{z_j}},
\end{equation}
given that the measurements are in the standard basis $\{\ket{\vec{z_i}}\}_{i=1}^{2^n}$ such that $\vec{z}_i\in\{0,1\}^{\otimes n}$.

Thus, as the diagonal elements of $\rho$ have the relationships between them unchanged by dephasing noise acting on a single qubit, we see that the VLS algorithm has resilience to single-qubit dephasing noise occurring at the end of a circuit.
\end{specialproof}

We can now extend this argument to multiple qubits, by noting that the application of dephasing noise to each qubit at the end of a quantum circuit would have the form,

\begin{equation}\label{eq:Dephasing_Noise_Multi_Qubits}
    \varepsilon(\rho) = \varepsilon_n(\varepsilon_{n-1}(\ldots \varepsilon_1(\rho)\ldots )),
\end{equation}
where $\varepsilon_i$ is the noise applied to each of the $i$ qubits.
(Note that \textit{all} of the noise must still be assumed to occur at the end of the circuit.)
Then, $\varepsilon_1(\rho)$ has the form of the above corollary,
\begin{equation}\label{eq:Dephasing_Noise_Multi_Qubits_1}
\varepsilon_1(\rho) = p_1\rho + (1-p_1)Z\rho Z. 
\end{equation}
Applying $\varepsilon_2(\rho)$ to the result of eq.~\eqref{eq:Dephasing_Noise_Multi_Qubits_1} gives,
\begin{equation}\label{eq:Dephasing_Noise_Multi_Qubits_2}
\begin{split}
\varepsilon_2(\varepsilon_1(\rho)) & = p_2(p_1\rho + (1-p_1)Z\rho Z) \\
& + (1-p_2)Z(p_1\rho + (1-p_1)Z\rho Z)Z \\ 
& = p_1p_2\rho + p_2(1-p_1)Z\rho Z \\
& + p_1(1-p_2)Z\rho Z \\
& + (1-p_1)(1-p_2)Z^{\otimes 2}\rho Z^{\otimes 2}.
\end{split}
\end{equation}
Similarly, we find that $\varepsilon_3(\rho)$ has the simplified form,
\begin{equation}\label{eq:Dephasing_Noise_Multi_Qubits_3}
\begin{split}
\varepsilon_3(\varepsilon_2(\rho)) & = p_1p_2p_3\rho + p_2p_3(1-p_1)Z\rho Z \\
& + p_1p_3(1-p_2)Z \rho Z \\
& + p_3(1-p_1)(1-p_2)Z^{\otimes 2}\rho Z^{\otimes 2} \\
& + p_1p_2(1-p_3)Z\rho Z \\
& + p_2(1-p_1)(1-p_3)Z^{\otimes 2}\rho Z^{\otimes 2} \\
& + p_1(1-p_2)(1-p_3)Z^{\otimes 2}\rho Z^{\otimes 2} \\
& + (1-p_1)(1-p_2)(1-p_3)Z^{\otimes 3}\rho Z^{\otimes 3}.
\end{split}
\end{equation}

We thus see that $\varepsilon_n(\rho)$ will have the form,
\begin{equation}\label{eq:Dephasing_Noise_Multi_Qubits_General_Form}
\varepsilon_n(\rho) = \tilde{\rho} = \sum_\mu{q_\mu \boldsymbol{Z}_\mu \rho \boldsymbol{Z}_\mu},
\end{equation}
where $\boldsymbol{Z}_\mu = \{I, Z\}^{\otimes n}$ because $Z^{\otimes n}$ for even $n$ is $I$ and for odd $n$ is $Z$.
Furthermore, the postulates of quantum mechanics require that $\sum_\mu{q_\mu}=1$.

We can now consider eq.~\eqref{eq:Dephasing_Noise_Multi_Qubits_General_Form} in light of the solution stored by VLS.
We illustrated in proving  Corollary~\ref{th:Dephasing_Noise_Single_Qubit} that VLS stores information on the diagonals of the quantum state density operator it prepares.
This means that the solution is stored in the measurement values $p(\ket{z_i})$, so, to show that VLS exhibits resilience to dephasing noise applied to each qubit at the end of a circuit, we show that $p(\ket{z_i})$ is the same whether or not dephasing noise is applied to one or more qubits.

First, we note that $p(z_i)$ in the case of dephasing noise applied to a single qubit is defined as,
\begin{equation}\label{eq:Meas_Def_Dephasing_Noise_Single_Qubit}
p(z_i) = \bra{z_i}\rho\ket{z_i},
\end{equation}
because the single-qubit noise leaves $\rho$ unchanged.
Similarly, in the case of dephasing noise applied to multiple qubits, $p(z_i)$ is defined as,
\begin{equation}\label{eq:Meas_Def_Dephasing_Noise_Multi_Qubit}
p(z_i) = \bra{z_i}\varepsilon_n(\rho)\ket{z_i} = \bra{z_i}\tilde{\rho}\ket{z_i}.
\end{equation}
Plugging in the result of eq.~\eqref{eq:Dephasing_Noise_Multi_Qubits_General_Form} gives,
\begin{equation}\label{eq:Meas_Def_Dephasing_Noise_Multi_Qubit_Simplified}
\begin{split}
p(z_i) & = \bra{z_i}\sum_\mu{\boldsymbol{Z}_\mu\rho\boldsymbol{Z}_\mu q_\mu}\ket{z_i} \\
& = \sum_\mu{q_\mu\bra{z_i}\boldsymbol{Z}_\mu\rho\boldsymbol{Z}_\mu\ket{z_i}}. 
\end{split}
\end{equation}
Here, as in the proof of Corollary~\ref{th:Dephasing_Noise_Single_Qubit}, the dual $\boldsymbol{Z_\mu}$'s will cancel any effect on $\rho, \bra{z_i},$ and $\ket{z_i}$.
Furthermore, we stated that $\sum_\mu{q_\mu} = 1$, so eq.~\eqref{eq:Meas_Def_Dephasing_Noise_Multi_Qubit_Simplified} simplifies to,
\begin{equation}\label{eq:Meas_Def_Dephasing_Noise_Multi_Qubit_Final}
p(z_i) = \bra{z_i}\rho\ket{z_i}, 
\end{equation}
which is precisely what it was for the case of single-qubit dephasing noise.

Therefore, as the diagonal elements of $\rho$ have the relationships between them unchanged by dephasing noise applied at the end of a quantum circuit,the VLS algorithm is resilient to dephasing noise that occurs at the end of a circuit.
\end{specialproof}

\subsubsection{Global Depolarising Noise}
Global depolarising noise is caused by the tendency of the quantum state to shift towards a maximally-mixed state, which contains no relevant information beyond noise~\cite{nielsen2000quantum}.
The maximally-mixed state is mathematically represented by $\frac{I}{2^n}$, where $n$ is the number of qubits in the state.
We can represent global depolarising noise mathematically as,
\begin{equation}\label{eq:Depolorising_Noise_Definition}
\tilde{\rho} = \varepsilon_p(\rho) = p \rho + \frac{(1-p)}{2^n}I,
\end{equation}
where---as with dephasing noise---$p$ is the probability that the state remains unaffected by global depolarising noise, and $(1-p)$ is the probability that $\rho$ changes due to noise.

\begin{theorem}\label{th:Depolarising_Noise_Resilience}
The VLS algorithm is partially resilient to global depolarising noise that occurs at the end of each circuit layer, because global depolarising noise does not change the sign information of the diagonal elements of the density operator, $\rho$, representing the VLS algorithm's prepared state.
\end{theorem}

\begin{specialproof}
Any quantum circuit ansatz can be represented by a series of operations, each of which represents one of its unitary layers.
Recall that each layer is a series of gates with identical gate type but varying parameters that the VLS algorithm seeks to tune.
Let $U_l(\cdot)$ be one such ansatz layer of gates that run in parallel, and let $U(\boldsymbol{\theta})$ represent the 
entire ansatz, such that $U(\boldsymbol{\theta})=\prod_{l=1}^L U_l(\boldsymbol{\theta})$, where $L$ is the number of layers of gates, which scales linearly with the number of ansatz layers.
So, the quantum state after each layer, $l=1,...,L$ can be represented as,
\begin{equation}\label{eq:State_After_Layer}
\rho_{l}= U_l \rho_{l-1} U_l^{\dagger},
\end{equation}
with $\rho_{l-1}$ being the state immediately before application of layer $l$ gates, $\rho_0 = \ket{0}^{\otimes n}\bra{0}^{\otimes n}$ being the initial state,
and $\rho_L$ being the final, resulting state.

Assume that global depolarising noise occurs after each layer. In other words, the quantum state $\tilde{\rho}_l$ after each layer and noise is
\begin{equation}\label{eq:State_After_Layer_Noise}
\tilde{\rho}_{l} = \varepsilon_p(\rho_l) = \varepsilon_p\left(U_l \rho_{l-1} U_l^{\dagger}\right).
\end{equation}
Note that the initial state remains $\rho_0$ because we assume that no global depolarisation occurs until after the first ansatz layer (i.e., $U_1$).

\begin{corollary}\label{th:Depolarising_Noise_Form}
Global depolarising noise scales exponentially with the number of layers while keeping its form
$$
\tilde{\rho}_l = p^{l} \rho_{l} + \frac{1-p^l}{2^n}I.
$$
\end{corollary}

\begin{specialproof}
We can prove that global depolarising noise keeps the form given in Corollary~\ref{th:Depolarising_Noise_Form} by induction.
First, the base case $l=0$ is true because---as stated above---the system has $\rho_0$ in both the situations of no global depolarising noise and global depolarising noise after each layer.
And, $\tilde{\rho_0} = p^0\rho_0 + \frac{(1-p^0)}{2^n}I = \rho_0$, which shows that $\tilde{\rho_0}$ can be correctly described using the form in Corollary~\ref{th:Depolarising_Noise_Form}.

So, we need next show the inductive step, namely that,
\begin{equation}\label{eq:Inductive_Step}
\tilde{\rho}_{l} = \varepsilon_p(U_l \tilde{\rho}_{l-1} U_l^{\dagger}) = p^{l} \rho_{l} + \frac{(1-p^l)}{2^n}I,
\end{equation}
for all $l=1,..,L$.
Assume that the system after $1 \le k-1 \le L$ layers and $k-1$ applications of global depolarising noise is given by
\begin{equation}\label{eq:Inductive_Hypothesis}
\tilde{\rho}_{k-1} = p^{k-1}\rho_{k-1} + \frac{1-p^{k-1}}{2^n}I.
\end{equation}
Then, applying the next layer of gates and noise provides state
\begin{equation}
\tilde{\rho}_k = \varepsilon_p(U_k \tilde{\rho}_{k-1} U_k^\dagger).
\end{equation}
Simplifying by plugging in eq.~\eqref{eq:Inductive_Hypothesis} for $\tilde{\rho}_{k-1}$ gives,
\begin{equation}
\begin{split}
\tilde{\rho}_k & = \varepsilon_p(U_k(p^{k-1}\rho_{k-1} + \frac{1-p^{k-1}}{2^n}I)U_k^\dagger) \\ 
& = \varepsilon_p(p^{k-1}(U_k\rho_{k-1}U_k^\dagger) + \frac{1-p^{k-1}}{2^n}U_k I U_k^\dagger) \\ 
& = \varepsilon_p(p^{k-1}U_k\rho_{k-1}U_k^\dagger + \frac{1-p^{k-1}}{2^n}I).
\end{split}
\end{equation}

Applying the definition of $\varepsilon_p$ from eq.~\eqref{eq:Depolorising_Noise_Definition} gives,
\begin{gather*}
\tilde{\rho}_k = p(p^{k-1}U_k\rho_{k-1}U_k^\dagger + \frac{1-p^{k-1}}{2^n}I) + \frac{(1-p)}{2^n}I \\ 
= p^k U_k\rho_{k-1}U_k^\dagger + \frac{p(1-p^{k-1})}{2^n}I \\
+ \frac{(1-p)}{2^n}I \\ 
= p^k U_k\rho_{k-1}U_k^\dagger + \frac{(1-p^k)}{2^n}I.
\end{gather*}

Finally, noting that eq.~\eqref{eq:State_After_Layer_Noise} is present in the above, we can simplify to,
\begin{gather*}
\tilde{\rho}_k = p^k \rho_{k} + \frac{(1-p^k)}{2^n}I,
\end{gather*}
which is the form we sought to show for $1 \le k \le L$.
Thus, Corollary~\ref{th:Depolarising_Noise_Form} is proven.
\end{specialproof}

Having established the effect of global depolarising noise after each layer, we need only to establish that the values on the diagonal of the resulting density operator after layer $i$ have the same sign as those in the non-noise-affected density operator after layer $i$ (i.e., we need to compare elements on the diagonal of $\tilde{\rho_l}$ and $\rho_l$).
Consider elements at indices $aa$ and $bb$ in $\tilde{\rho_l}$ such that $0 \le l \le K$, and $aa, bb \in \{00, 11, ..., (2^n-1)(2^n-1)\}$.
Then, by the result of Corollary~\ref{th:Depolarising_Noise_Form},
\begin{equation}\label{eq:Values_of_Depolarising_DO}
\begin{split}
    \tilde{\rho_l}_{aa} & = p^l{\rho_l}_{aa} + \frac{1-p^l}{2^n}, \\
    \tilde{\rho_l}_{bb} & = p^l{\rho_l}_{bb} + \frac{1-p^l}{2^n}.
\end{split}
\end{equation}

We seek to show that the elements along the diagonal of $\tilde{\rho_l}$ have the same signs as the diagonal elements in $\rho_l$; this indicates that the sign information is preserved in the presence of global depolarising noise.
We can illustrate this by considering the subtraction of $\tilde{\rho_l}_{bb}$ from $\tilde{\rho_l}_{aa}$, which gives,
\begin{equation}\label{eq:Subtracing_DO_Values}
\begin{split}
   & \tilde{\rho_l}_{bb} - \tilde{\rho_l}_{aa}
    = p^l{\rho_l}_{aa} +  \frac{1-p^l}{2^n} \\
    & - p^l{\rho_l}_{bb} - \frac{1-p^l}{2^n} \\
   & = p^l({\rho_l}_{aa} - {\rho_l}_{bb}),
\end{split}
\end{equation}
for all $l=0,...,L$.
Because $p^l$ is a probability value that is thus always positive, the above illustrates that the sign information on the diagonal elements of $\tilde{\rho}_l$ is the same as that on the diagonal elements of $\rho_l$.
And finally, because the VLS algorithm stores the information we seek on the diagonal elements of the density operator (as proven for Theorem~\ref{th:Dephasing_Noise_Resilience}), then we see that the VLS algorithm is partially resilient to global depolarising noise because the sign of the information contributing to the solution is preserved in the presence of global depolarising noise after each layer of the circuit.
\end{specialproof}

\subsection{`Smart Encoding' to Reduce Complexity of Obtaining Solution}\label{sec:Smart_Encoding}
As discussed in the Introduction, quantum computers provide efficiency gains in \textit{computing}, but not necessarily \textit{obtaining}, problem solutions.
We can motivate this with a straightforward example; consider the fracture network of approximately $10^{15}$ nodes that we considered at the beginning of the paper.
A quantum computer will store the solution vector to a linear system using only about 50 qubits.
However, those 50 qubits will represent a solution vector of approximately $2^{50} \approx 10^{15}$ elements.
Furthermore, each of the elements in this desired solution vector is a probability, meaning a value between 0 and 1.
So, our sought solution is $10^{15}$ values between 0 and 1.

When we take measurements from a quantum computer, we obtain individual solutions that allow us to form a probability distribution from which we can obtain the desired vector of probabilities.
So, in a situation such as the $10^{15}$ nodes example, we can thus see that the number of measurements necessary to establish a probability distribution for an exponentially-increasing number of nodes between 0 and 1 is also going to grow exponentially: even if we needed only one measurement per node, we would require $10^{15}$ measurements.

Mathematically, we can show this as follows.
Consider the probability amplitudes, $|c_i|^2$ for each of the $i$ elements in the vector of solutions ($i=0, ..., n-1$ for $n$ nodes).
These must sum to one, giving us that,
\begin{equation}\label{eq:Probs_Sum_to_1}
\sum_i{|c_i|^2}=1.
\end{equation}
Therefore,
\begin{equation}\label{eq:Each_c_i}
|c_i| \propto \frac{1}{\sqrt{n}},
\end{equation}
because there are $n$ nodes.
Therefore, each probability amplitude is exponentially small.
Furthermore, taking $N$ measurements, or `shots,' definitionally provides a solution precision of $\frac{1}{\sqrt{N}}$.
Therefore, to make $\frac{1}{\sqrt{N}}$ proportional to $\frac{1}{\sqrt{n}}$, $N$ must be proportional to $n$, meaning that the number of shots we need grows exponentially.

\begin{figure}[t]
\centerline{\includegraphics[width=\linewidth]{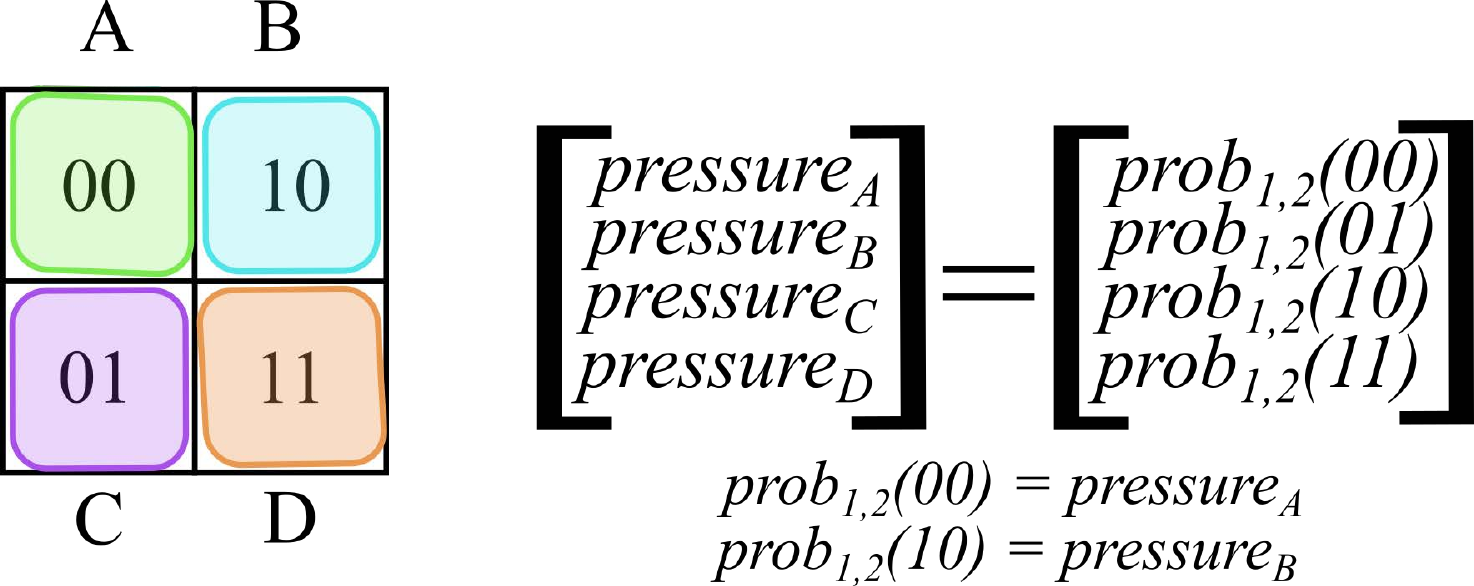}}
\caption{\textbf{A $2\times 2$ fracture network with measured pressures for each node and no `smart encoding.'}  Each of the nodes is labelled with a letter (A through D), and contains the index of the solution vector that will contain the pressure at that node.  The vectors illustrate the probabilities that provide each of the node pressures, and the equations below provide two specific examples.  First, the probability of measuring the first qubit to be 0 and the second qubit to be 0 provides the pressure solution for node A.  Similarly, the probability of measuring the first qubit to be 1 and the second qubit to be 0 provides the pressure solution for node B.}
\label{fig:Four_Node_No_Smart_Encoding}
\end{figure}

In this paper, we sought to assess the performance of VLS, meaning we needed to obtain every element of the solution, meaning every probability amplitude, to see how well VLS computed each element.
For this reason, we accepted the exponential cost.
Therefore, we obtained information as shown for a simple example in Fig.~\ref{fig:Four_Node_No_Smart_Encoding}.
The grid represents the nodes in the fracture region, each of which is denoted with a letter (A through D).
Each of the numbers in the nodes represents the index (in binary notation) of the element in the probability solution vector corresponding to the normalized pressure solution for that node.
As shown in the vectors in Fig,~\ref{fig:Four_Node_No_Smart_Encoding}, the number in each node is \textit{also} the state (for each of the qubits, 1 and 2) whose probability is associated with the pressure for the desired node.
So, measuring both of the qubits a number of times allows us to craft a probability distribution with one state probability per node.

While our approach was permissible for benchmarking the performance of the VLS algorithm, it is not sustainable for large fracture flow problems. 
Using exponential resources to obtain the solution would negate the advantage of quantum computing's speedup, and would become prohibitive for sufficiently large problems.
Fortunately, the nature of fracture flow problems is such that one often desires the solution at only a subset of nodes.
For situations in which this is not the case, however, we present `smart encoding' as an alternative way to obtain aggregated information about more than one node without exponential complexity.

\begin{figure}[t]
\centerline{\includegraphics[width=\linewidth]{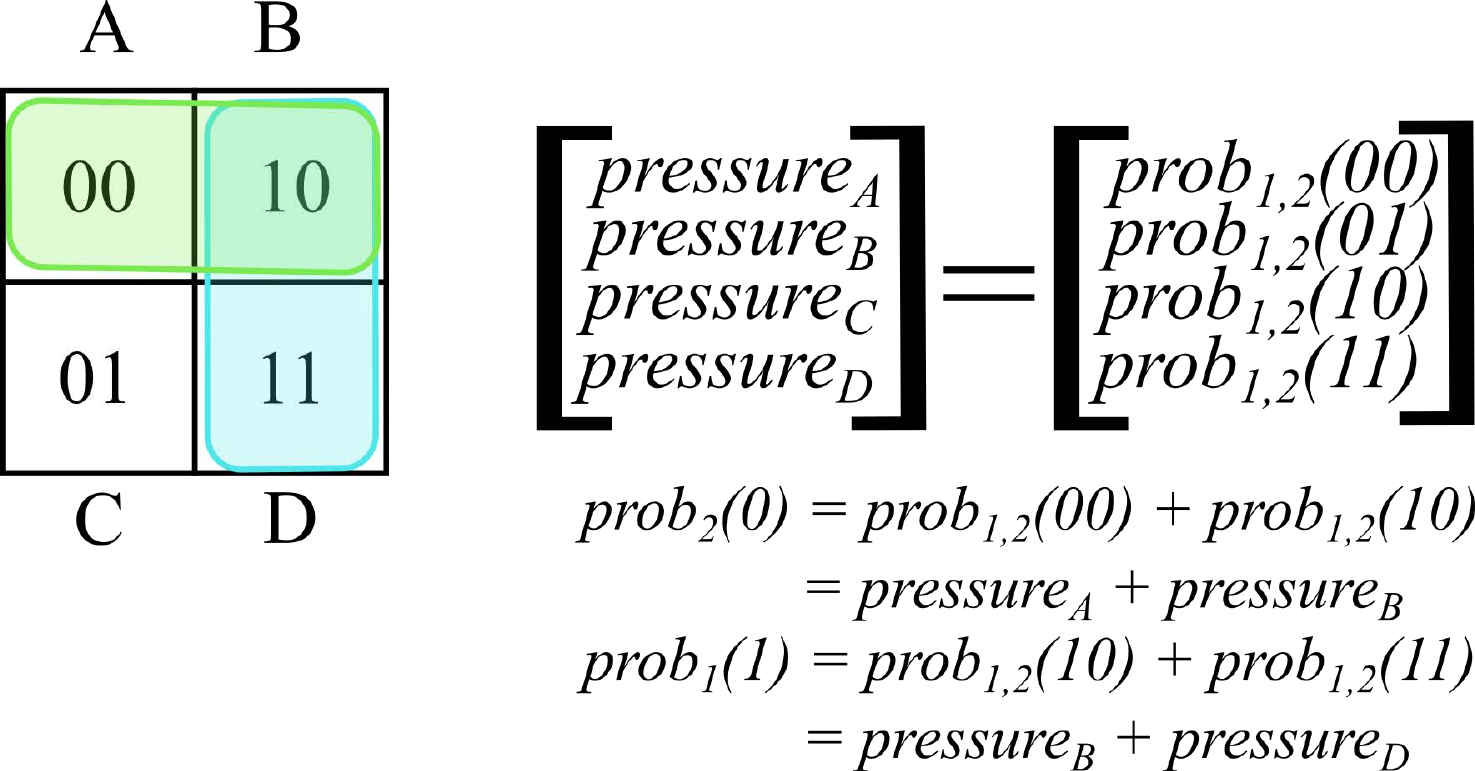}}
\caption{\textbf{A $2\times 2$ fracture network using `smart encoding.'}  Again, each node is labelled with a letter (A through D), and contains the index of the solution vector that will contain the pressure at that node.  While the vectors illustrating the probabilities that equate to each pressure are the same, the equations illustrate how we can measure just one of the qubits to obtain information about multiple nodes.  First, the probability of measuring the first qubit to be 0 and the second qubit to be 0 provides the pressure solution for node A.  Similarly, the probability of measuring the first qubit to be 1 and the second qubit to be 0 provides the pressure solution for node B.}
\label{fig:Four_Node_Smart_Encoding}
\end{figure}

Consider Fig.~\ref{fig:Four_Node_Smart_Encoding}, which presents a `smart encoded' use of the situation in Fig.~\ref{fig:Four_Node_No_Smart_Encoding}.
Here, we can obtain information about a row or column of nodes by measuring just a single qubit, instead of measuring both.

\begin{figure}[t]
\centerline{\includegraphics[width=\linewidth]{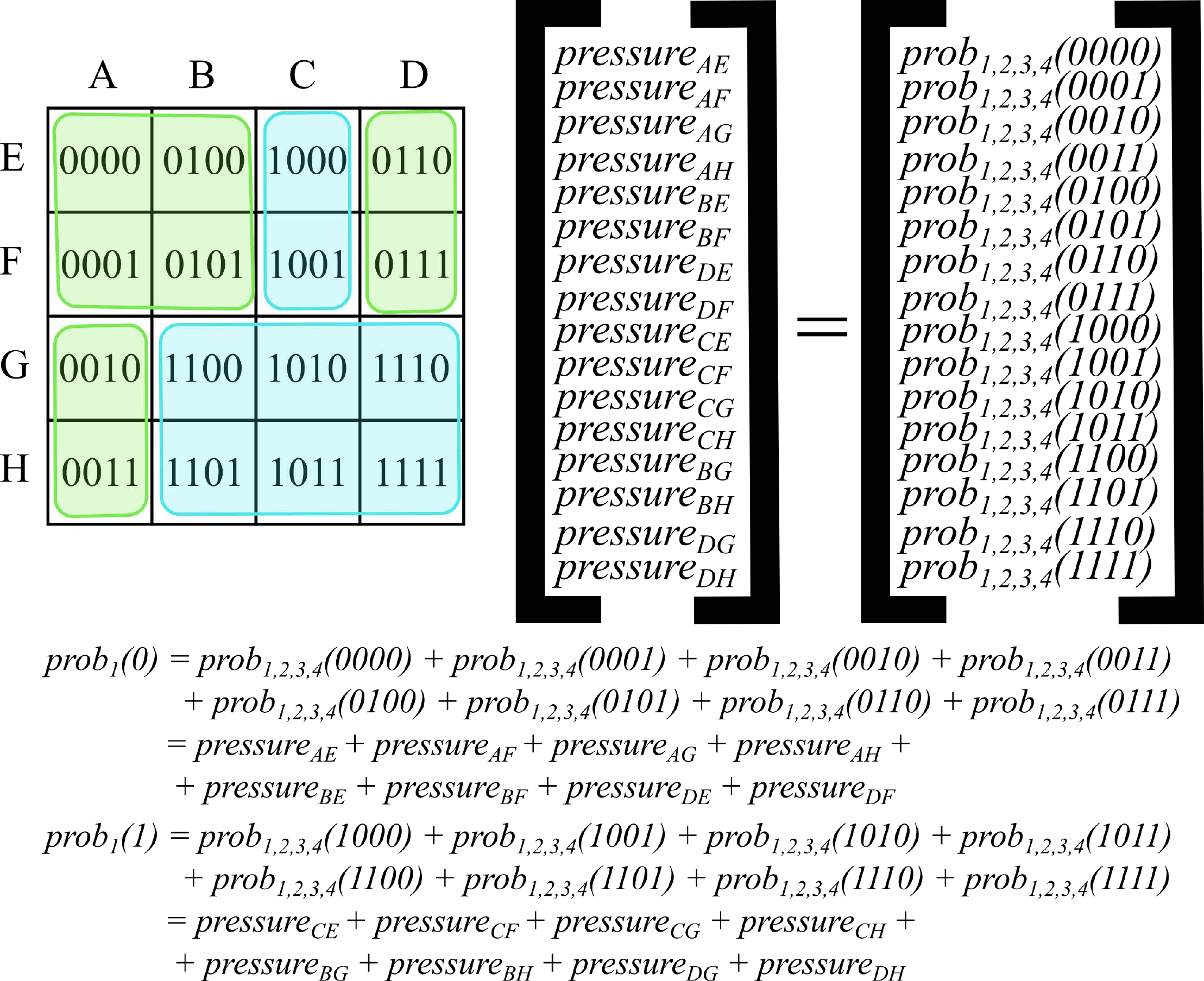}}
\caption{\textbf{A $4\times 4$ pitchfork fracture network using `smart encoding.'}  This example illustrates how arranging the indices---and thus solution probabilities---in problem-specific ways can reduce the measurements required to obtain information about more than one node.  Here, only the first qubit need be measured to differentiate the pressure throughout the pitchfork fracture from that in surrounding region.}
\label{fig:Pitchfork_Smart_Encoding}
\end{figure}

While the benefits of smart encoding are limited in Fig.~\ref{fig:Four_Node_Smart_Encoding}'s trivial example, they become significantly more pronounced for larger problems.
For example, consider larger regions in which information might be sought about certain pieces of that network.
The desired solution could be the pressure in a \textit{set} of nodes comprising a fracture, such as in Fig.~\ref{fig:Pitchfork_Smart_Encoding}, which presents a 16-node problem containing a pitchfork fracture.
Here, if we arrange the indices in a thoughtful way, we can obtain information about the entire network when measuring only the first qubit.
By determining the probability that the first qubit is zero, we learn the pressure throughout the non-pitchfork fracture region.
Subtracting that value from one then provides the probability that the first qubit is one, which consequently provides the pressure throughout the pitchfork fracture.
Future work should more fully consider both the benefits of and automated procedures for reformulating problems using smart encoding.

\subsection{Preliminary Results with Varying Permeability}\label{sec:varying_permeability}
After successfully scaling the 6x8 problem to larger region sizes, we returned to the smallest pitchfork problem to consider the affects of varying permeability.
We once again embedded a pitchfork fracture in a region of lower permeability, but this time, the pitchfork contained two permeabilities.
Specifically, the right-most branch of the pitchfork had a permeability that was either 10 times, 100 times, 1000 times, or 10,000 times larger than the rest of the pitchfork.
While the resulting system has the same size as the uniform-permeability pitchfork with five qubits, it is more complex because its elements vary by orders of magnitude more than in the uniform-permeability case.

Using the same general approach described in Sec.~\ref{sec:VLS_methods}, we obtained parameters for a parameterized quantum circuit that we then ran on the \texttt{ibmq\_mumbai} machine.
The training involved one specific procedural difference.
Instead of training with forty instances of randomly-initialized parameters, we began with a set of  trained parameters from earlier situations.
For example, when training the circuit for a situation with a right-branch permeability of 10 times greater than the rest of the pitchfork, the initial parameters were those trained for the uniform-permeability pitchfork.
And when training the circuit for a situation with a right-branch permeability of 100 times greater than the rest of the pitchfork, the initial parameters were either those trained for the uniform-permeability pitchfork or those trained for the right-branch-10-times-greater pitchfork.

We found that it was more difficult to obtain parameters that provided a high-fidelity result for these varying permeability problems than it was for the uniform-permeability variants.
When running the best trained circuits on a classical simulator absent hardware noise, we achieved an average fidelity of 0.9555, to four digits of precision, illustrating that the classical optimization process had a more difficult time obtaining parameters than in the uniform-permeability-pitchfork case.
Indeed, the fidelities suggest that each 10 times increase in permeability made optimal parameters harder for the classical optimization process to find.
Specifically, although we were able to find parameters with a maximum 0.9808 fidelity for the situation with a rightmost-branch permeability of 10 times greater than the rest of the pitchfork, we found parameters with maximum fidelities of only 0.9763, 0.9757, and 0.9756 for problems with rightmost-branch permeabilities of 100 times, 1000 times, and 10,000 times greater than the rest of the pitchfork, respectively.

\begin{figure}[t]
\centerline{\includegraphics[width=\linewidth]{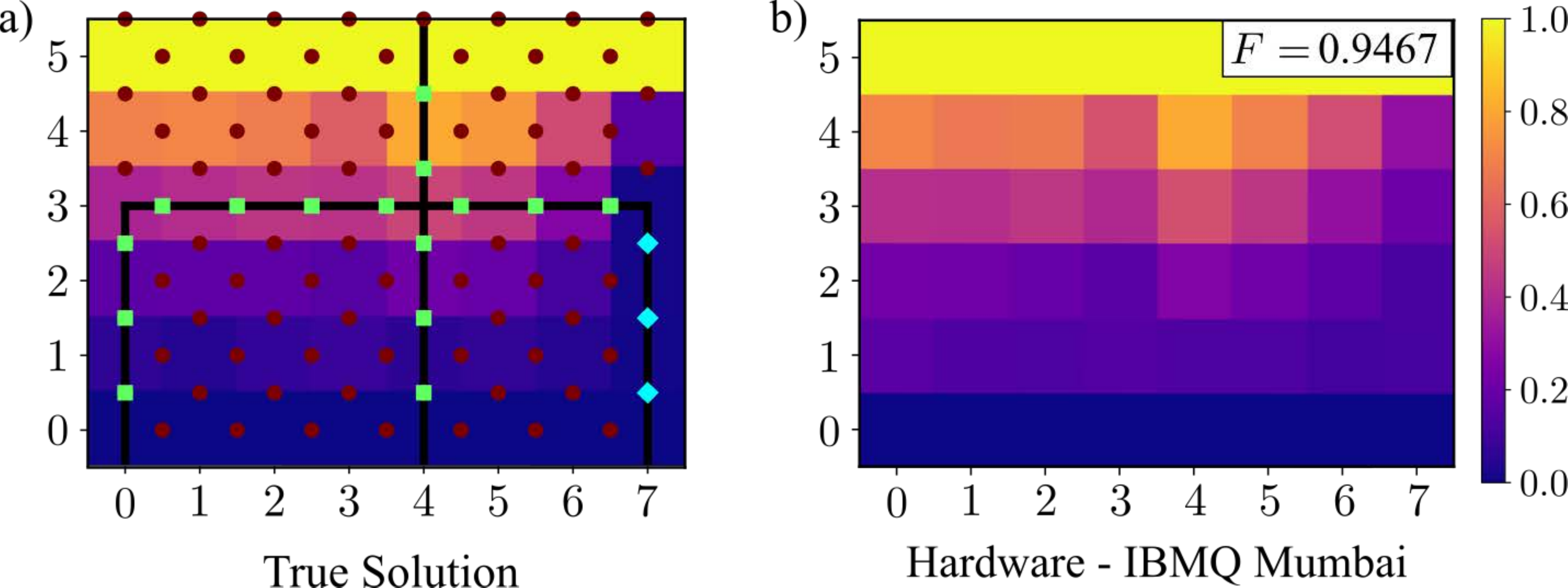}}
\caption{\textbf{Solving a 6x8 Problem with Varying Pitchfork Permeability.}  Subfigure (a) illustrates the known, classically computed solution with overlaid permeabilities.  The blue diamonds are a permeability 10,000 times greater than that of the remainder of the pitchfork fracture, represented with green squares.  The inner $4 \times 8$ nodes are the sought-after pressure values.  Subfigure (b) is the solution from quantum hardware (specifically, qubits 0, 1, 4, 7, and 10 of the \texttt{ibmq\_mumbai} machine).  This solution has fidelity 0.9467, to four figures; the trained circuit achieved a maximum fidelity of 0.9756 when run on a noiseless classical simulator.}
\label{fig:Varying_Perm_Results}
\end{figure}

As expected, when we ran the circuits on the quantum hardware, noise reduced the fidelities achieved during training.
Specifically, we achieved maximum fidelities of 0.9651, 0.9432, 0.9469, and 0.9467 for problems with rightmost-branch permeabilities of 10 times, 100 times, 1000 times, and 10,000 times greater than the rest of the pitchfork, respectively.
Again, we used the \texttt{ibmq\_mumbai} machine (qubits 0, 1, 4, 7, and 10) with 12 runs of 8192 shots each for a total of 98,304 shots.
Fig.~\ref{fig:Varying_Perm_Results} illustrates results from the 10,000 times greater case.

While all of our experiments achieved fidelities greater than 0.9, the dip in accuracy demonstrates that there is room for improvement when it comes to solving more complex fracture problems with quantum algorithms.
Fortunately, there are a number of forms such improvement could take, and here, we briefly discuss three.

First, we could further optimize the \textit{classical} portion of the VLS algorithm.
Classical optimization is complex in its own right, and we did not significantly tune the classical optimization process for the varying-permeability problems.
The increased complexity of these problems suggests that such tuning might be worthwhile.
Specifically, experimenting with which classical optimizer to apply~\cite{pellow2021comparison} and how many layers of the ansatz to use might offer significant improvement, instead of applying the same choices that worked for uniform-permeability pitchfork cases.
These considerations in particular might be especially fruitful because the final cost values for the trained circuits suggested that the optimizer was becoming stuck in a local minimum; alternative optimization methods specifically designed with that in mind might thus address this shortcoming.

Second, we could apply preconditioning methods to lower the condition-number of the matrix in the LSP prior to solving~\cite{golden2022quantum}.
Ref.~\cite{golden2022quantum} suggests that reducing the condition number by applying methods specifically designed for solving LSPs with quantum algorithms can make the problem significantly less complex---and thus significantly less prone to error.

Third and finally, we might apply quantum error mitigation techniques.
Quantum error mitigation is a relatively new field that is designed to address the shortcomings of near-term quantum hardware.
Like variational algorithms, quantum error mitigation algorithms are designed for the NISQ hardware that we have today, and thus seek not to \textit{correct} errors, but instead to \textit{work around} them, in many cases by actually using the noise present, instead of solely trying to eliminate it~\cite{mitiq-qem}.
For example, some quantum error mitigation approaches attempt to add selected gates to the circuit that will increase the amount of noise in such a way that the \textit{net} amount of noise is reduced, due to interference between the `automatically-present' noise and that added~\cite{mitiq-qem}.
Another method known as zero-noise extrapolation supposes a given amount of noise present, increases that supposed amount of noise by a known factor greater than one, and then uses results with both noise levels to extrapolate what the results would be with a noise level of zero.~\cite{mitiq-qem,temme2017error}
In Ref.~\cite{o2022near}, zero-noise extrapolation contributed to significantly reducing error---in some cases by an order of magnitude---suggesting that the technique might provide at least some benefit were we to try it here.

Thus, our preliminary results for geologic situations involving multiple permeabilities within the pitchfork suggest that there is room for refinement when it comes to solving such problems with quantum algorithms and, specifically, with VLS.
Fortunately, there are at least the three above readily-available avenues for such improvement.

\end{document}